\documentclass[preprint,review, 12pt]{elsarticle}

\usepackage{amssymb}
\usepackage{longtable}
\usepackage[normalem]{ulem}
\usepackage{amsmath}
\usepackage{caption}
\usepackage{subcaption}
\usepackage{graphicx}
\usepackage{xcolor}
\usepackage[english]{babel}
\usepackage{longtable}

\def\R{\mathbb{R}}
\def\Joule{\mathrm{J}}
\def\kilogram{\mathrm{kg}}
\def\Kelvin{\mathrm{K}}
\def\meter{\mathrm{m}}
\def\sec{\mathrm{s}}
\def\Watt{\mathrm{W}}
\def\mole{\mathrm{mole}}
\def\Newton{\mathrm{N}}

\newcommand{\new}[1]{#1}
\journal{International Journal of Heat and Mass Transfer}

\begin{document}

\begin{frontmatter}

%% Title, authors and addresses

%% use the tnoteref command within \title for footnotes;
%% use the tnotetext command for theassociated footnote;
%% use the fnref command within \author or \affiliation for footnotes;
%% use the fntext command for theassociated footnote;
%% use the corref command within \author for corresponding author footnotes;
%% use the cortext command for theassociated footnote;
%% use the ead command for the email address,
%% and the form \ead[url] for the home page:
%% \title{Title\tnoteref{label1}}
%% \tnotetext[label1]{}
%% \author{Name\corref{cor1}\fnref{label2}}
%% \ead{email address}
%% \ead[url]{home page}
%% \fntext[label2]{}
%% \cortext[cor1]{}
%% \affiliation{organization={},
%%             addressline={},
%%             city={},
%%             postcode={},
%%             state={},
%%             country={}}
%% \fntext[label3]{}

\title{Evaporative Refrigeration Effect in Evaporation and Condensation between Two Parallel Plates}

%% use optional labels to link authors explicitly to addresses:
%% \author[label1,label2]{}
%% \affiliation[label1]{organization={},
%%             addressline={},
%%             city={},
%%             postcode={},
%%             state={},
%%             country={}}
%%
%% \affiliation[label2]{organization={},
%%             addressline={},
%%             city={},
%%             postcode={},
%%             state={},
%%             country={}}

\author[PC]{Peiyi Chen\corref{cor1}} %% Author name
\ead{pchen345@wisc.edu}
%% Author affiliation
\cortext[cor1]{Corresponding author.}
\affiliation[PC]{organization={Department of Mathematics, University of Wisconsin-Madison},%Department and Organization
            addressline={480 Lincoln Drive}, 
            city={Madison},
            postcode={53706}, 
            state={WI},
            country={USA}}

\author[QL]{Qin Li} %% Author name
\ead{qinli@math.wisc.edu}
%% Author affiliation
\affiliation[QL]{organization={Department of Mathematics, University of Wisconsin-Madison},%Department and Organization
            addressline={480 Lincoln Drive}, 
            city={Madison},
            postcode={53706}, 
            state={WI},
            country={USA}}
            
\author[GC]{Gang Chen} %% Author name
\ead{gchen2@mit.edu}
\affiliation[GC]{organization={Department of Mechanical Engineering, Massachusetts Institute of Technology},%Department and Organization
            addressline={77 Massachusetts Avenue}, 
            city={Cambridge},
            postcode={02139}, 
            state={MA},
            country={USA}}          
%% Abstract
\begin{abstract}
%% Text of abstract
It is well-known that evaporation can lead to cooling.  However, little is known that evaporation can actually create a refrigeration effect, i.e., the vapor phase temperature can drop below the temperature of the cooling wall.
This possibility was recently pointed out via modeling based on a n approximate quasi-continuum approach.  This work examines this effect rigorously by studying evaporation and condensation between two parallel plates by coupling the solution of the Boltzmann transport equation in the vapor phase with the continuum treatments in both liquid films.  Numerical results show that the vapor phase temperature at the evaporating side can be much lower than the coldest wall temperature at the condensing surface, \new{reaffirming the evaporative refrigeration effect.} \new{The present work further reveals that this effect is} caused by two mechanisms. \new{While the dominant mechanism is} the asymmetry in the \new{molecular} distribution between the outgoing and the incoming molecules \new{at the interface,} additional cooling occurs within the Knudsen layer due to the sudden expansion, similar to the Joule-Thomson effect, although with subtle differences in that the interfacial expansion is not an isenthalpic process. \new{The impacts of key parameters, including liquid thickness, Knudsen number, and accommodation coefficient, are investigated. The numerical simulation shows that with a thicker vapor, a thinner liquid film, and a larger accommodation coefficient, leads to stronger evaporative refrigeration effect.} \new{This} work will motivate future experiments to further confirm this prediction and explore its potential applications in air-conditioning\new{,} refrigeration\new{, and membrane distillation}.
\end{abstract}

%%Graphical abstract
%\begin{graphicalabstract}
%\includegraphics{grabs}
%\end{graphicalabstract}

%%Research highlights
% %
% The refrigeration effect can be exploited for potential air conditioning applications
%\end{highlights}

%% Keywords
\begin{keyword}
Refrigeration effect \sep Evaporation \sep Condensation \sep Boltzmann transport equation
%% keywords here, in the form: keyword \sep keyword

%% PACS codes here, in the form: \PACS code \sep code

%% MSC codes here, in the form: \MSC code \sep code
%% or \MSC[2008] code \sep code (2000 is the default)

\end{keyword}

\end{frontmatter}

%% Add \usepackage{lineno} before \begin{document} and uncomment 
%% following line to enable line numbers
%% \linenumbers

%% main text
%%

%% Use \section commands to start a section
\clearpage
\begin{longtable}[tbhp]{c l }%% placement specifier
%\centering
%\begin{tabular}{c l }%% Table column specifiers
  \multicolumn{2}{l}{\textbf{Nomenclature}} \\% & Description & Units \\ 
  \hline
  $c_p$ & specific heat, $\Joule\cdot\kilogram^{-1}\cdot\Kelvin^{-1}$\\
  $d_{0}$ & liquid thickness on the hot side, $\meter$ \\
  $d_{1}$ & liquid thickness on the cold side, $\meter$ \\
  $e$ & internal energy, $\Joule$\\
  $f$ & three-dimensional molecular distribution function,  $\kilogram\cdot\meter^{-3}\cdot\meter^{-3}\cdot\sec^{3}$ \\
  $g$ & one-dimensional molecular distribution function, $\kilogram\cdot\meter^{-3}\cdot\meter^{-1}\cdot\sec$ \\
  $h$ & one-dimensional energy distribution function,  $\kilogram\cdot\meter^{-3}\cdot\meter\cdot\sec^{-1}$ \\
  $\mathsf{h}$ & enthalpy,  $\Joule$ \\
  $j$ & energy flux,  $\Watt\cdot\meter^{-2} $ \\
  $k_w$ & liquid thermal conductivity,  $\Watt\cdot \meter^{-1}\cdot \Kelvin^{-1}$ \\
  $Kn$ & Knudsen number \\
  $m_0$ & molecular mass,  $\kilogram\cdot\mole^{-1}$ \\
  $\dot{m}$ & mass flux,  $\kilogram\cdot\meter^{-2} \cdot\sec^{-1}$\\
  $p$ & pressure,  $\Newton\cdot\meter^{-2} $\\
  $q$ & heat flux,  $\Watt\cdot\meter^{-2} $ \\
  $R$ & ideal gas constant,  $\Joule\cdot\kilogram\cdot\Kelvin^{-1}$\\
  $T$ & temperature, $\Kelvin$ \\
  $u$ & average velocity, $\meter\cdot\sec^{-1}$ \\
  $v$ & molecular velocity, $\meter\cdot\sec^{-1}$ \\
  $x$ & Cartesian coordinate,  $\meter$\\
  \hline
  
  \multicolumn{2}{l}{\textbf{Greek symbols}}\\
  \hline
  $\alpha$ & accommodation coefficient \\
  $\Gamma$ & latent heat, $\Joule\cdot\mathrm{kg}^{-1}$ \\
%\end{table}

%\begin{table}[tbhp]%% placement specifier
%\centering
%\begin{tabular}{c l }
  $\theta$ & $k_B T/m_0, \,\meter^2\cdot\sec^{-2}$\\
  $\rho$ & vapor density,  $\kilogram\cdot\meter^{-3}$ \\
  $\tau$ & relaxation time,  $\sec$\\
  $\phi$ & Maxwellian velocity distribution,  $\kilogram\cdot\meter^{-3}\cdot\meter^{-3}\cdot\sec^{3}$ \\
  \hline
  \multicolumn{2}{l}{\textbf{Subscripts}}\\
  \hline
  $c$ & characteristic \\
  $e$ & energy \\
  $i$ & $i=0$ for hot interface, and $i=1$ for cold interface \\
  $l_i$ & liquid side of the interface, $i=0$ for evaporating film, $i=1$ for condensing film \\
  $p$ & constant pressure \\
  $s$ & saturated state \\
  $v_i$ & vapor side of the interface, $i=0$ for evaporating film, $i=i$ for condensing film \\
  $w_i$ & wall, $i=0$ for hot wall, and $i=1$ for cold wall  \\
  \hline

  \multicolumn{2}{l}{\textbf{Nondimensional variables}}\\
  %\textbf{Nondimensional}\textbf{variables} \\
  \hline
  \new{$\hat{v}$} & \new{nondimensional molecular velocity} \\
  \new{$\hat{\theta}$} & \new{nondimensional temperature}  \\
  \hline
%\end{tabular}
%\caption{\new{Nomenclature.}}
\label{tbl1:variable_table}
%\end{tabular}
%\caption{\new{Nomenclature cont.}}
%\label{tbl1:variable_table2}
\end{longtable}

\clearpage

\section{Introduction}
In the Joule-Thomson effect, the sudden expansion of a fluid can \new{cause} the fluid temperature at the outlet \new{to drop} below that at the inlet, which is the underlying principle for most air-conditioning and refrigeration systems.  Evaporation of a fluid can also lead to cooling, i.e., the evaporative cooling, which is used in some air-conditioning systems~\cite{AB-thermal-2016, Moran-energy-1989}.  Putting a cup of water in \new{the} air, evaporation cools down the liquid and the surrounding air.  In this case, temperature of the surrounding is higher than \new{the} liquid and heat is transferred from the surrounding to water, i.e., the temperature is inverted.  Despite its familiarity, the unconventional temperature distribution can cause confusion: evaporation seems to carry heat away from the liquid while surrounding air transfers heat to the liquid.  Although this superficial contradiction to the second law of thermodynamics -- heat always spontaneously flows from hot to cold regions -- can be easily resolved by realizing that \new{in addition to temperature difference,} evaporation is \new{also} driven by the chemical potential difference between liquid water and water vapor in air, it does exemplify the complexity with evaporation.  In this work, another little-known consequence of evaporation \new{is discussed: that} the evaporating vapor can cool below the liquid temperature, \new{which we will call the evaporative refrigeration effect}.    

The evaporative refrigeration effect is related to temperature and density discontinuities at an evaporating interface.  Such discontinuities were anticipated by pioneering work of Hertz, Knudsen, and Schrage~\cite{Hertz-1882, Knudsen-1915, Schrage-1953, Carey-2020-book}. 
In 1971, Pao~\cite{Pao-1971-jump} predicted the existence of these discontinuities at an evaporating surface, starting from the Boltzmann transport equation \new{(BTE)}.  In the same year, \new{Pao} also investigated the parallel plate problem: evaporation from one surface and condensation on the other~\cite{Pao-oppo-sign-1971}.  Pao predicted that the vapor phase temperature profile is inverted: lower at the evaporation side and higher at the condensation side, opposing to the applied temperature difference between the hot evaporating wall and the cold condensing wall. \new{However,} Pao did not include any liquid film on either the evaporating or the condensing side. Instead, \new{Pao's studies} simply specified that the walls are at the saturation conditions.  Many subsequent studies have used the same assumptions to investigate both the single interface evaporation and the two parallel plate problems based on solving the Boltzmann transport equation numerically~\cite{Sone-1989, AM-1994-evaporation-numerics, CFF-1985-paradox, Cipolla-1974, Frezzotti-2007, ASY-1990-condensed-numerics, Koffman-1984, Shishkova-2016-interface, SHISHKOVA2017926, SHISHKOVA20199},  or via taking its moments~\cite{Ytrehus-1996, Aursand-2019, Vaartstra-2022}, Monte Carlo simulations~\cite{Jafari-2018}. \new{Although} molecular dynamics simulations~\cite{Frezzotti-2003-Physfluid, Meland-2004-PhysFL, Chandra-2020-JCP, BIRD-2022-molecular, Rokoni-2020-JCP, Zhakhovskii-1997, Zhakhovsky-2018, KON-2016, Meland-2003-PhysFluid, Tokunaga-2020-PhysFL} \new{can include the liquid region, the simulation domain size is limited}. Asymptotic solutions and discontinuous boundary conditions have also been derived~\cite{SO-1978-kinetic, Sone-Takata-Golse-2001, Barrett-1992-JC, Frezzotti-2011-PhysFL, Kucherov-1960, LABUNTSOV1979989} and used to couple to continuum treatment~\cite{Aoki-Cercignani-1983, Gatapova-2015-jump}. In a series of papers, Shishkova et al.~\cite{Shishkova-2016-interface, SHISHKOVA2017926, SHISHKOVA20199, LABUNTSOV1979989} worked on coupling BTE solution for the vapor phase with molecular dynamics simulation for the liquid phase and the interface. However, their studies are limited to van der Waals liquid. None of these studies has discussed possible refrigeration effect in the evaporation process in-depth, or the effect of parameter related to liquid films. 

Experimental measurements of the interfacial temperature discontinuities had also been reported, pioneered by Ward and co-workers~\cite{Fang-1999-PhysRevE, Ward-2001-interface, McGaughey-2002-ApplPhys, Badam-2007-experiment, Duan-2008-PhyE, Kazemi-Nobes-2017-Langmuir, Ghasemi-2011-JCP, Ghasemi-2010-PRL, Gatapova-2017-air-interface, Zhu-2013-Interfacial, Zhu-2009, Buffone-2004-IR, Kazemi-2017-Langmuir, Gatapova-2024-PhysFL, Popov-2005}, as well-summarized by Jafari et al.~\cite{Jafari-2019-JCP}. Most of these experiments reported that the vapor side has a higher temperature than that of the liquid surface, although a few experiments reported that the vapor side could actually be at a lower temperature than that of the liquid interface~\cite{Zhu-2013-Interfacial, Zhu-2009, Gatapova-2024-PhysFL}. These experiments cannot be modeled by directly solving the Boltzmann transport equation since the dimensions of the experiments are much larger than the vapor mean free path.  Besides, previous efforts in solving the Boltzmann transport equation had completely neglected the liquid layer. Approximate boundary conditions, such as the Hertz-Knudsen and the Schrage equations~\cite{Schrage-1953, Carey-2020-book} have too many variables that cannot be uniquely used to determine the interfacial temperature discontinuities.  Such difficulties have motivated alternative modeling approaches~\cite{Ward-1999-PhysRevE, Bedeaux-1990-Physica}. None of them predicted a lower temperature at the vapor side. \new{In addition, past experiments have focused on evaporation and condensation at a single interface.  Due to experimental challenges, the inverted vapor-phase temperature profile predicted for evaporation and condensation between two parallel plates has not been experimentally demonstrated yet since its first prediction by Pao~\cite{Pao-oppo-sign-1971} over 50 years ago.  Real experiments will necessarily include liquid regions on both the evaporating and condensation sides. The omission of these liquid regions in the past modeling work further hinders a full understanding of this classical problem, which is not only important from a fundamental perspective, but also for practical applications such as vapor chambers in microelectronics cooling and membrane distillation~\cite{Deshmukh_membrane_distillation_2018, WEIBEL_heat_flux_2013, Yan_Chen_condensation_2022}.}

Chen recently tackled the above difficulties by introducing an additional heat flux condition in addition to the often-used Schrage equation for the mass flux, such that the transport across the Knudsen layer can be approximated by density and temperature discontinuities~\cite{GC-2022-flux-bc, GC-2024-inversion}. With these boundary conditions, temperature distributions for both the liquid and vapor sides and temperature and density discontinuities at the interface can be obtained using continuum approximations for both the liquid and the vapor regions. Chen applied such quasi-continuum approach to treat evaporation from and condensation into a single liquid layer and showed that the experimental results in the past can be explained. The vapor side temperature can be lower or higher than that of the liquid surface temperature, depending on the \new{liquid} film thickness and the vapor side boundary conditions. The same approach was extended to treat the effect when noncondensable gas is present~\cite{GC-2025}.  For the two parallel plate problem, \new{Chen~\cite{Chen-2023-parallel} applied the heat flux and mass flux conditions and} showed temperature inversion as originally predicted by Pao~\cite{Pao-oppo-sign-1971}.  Even more interestingly, the modeling result shows that the vapor region could be significantly colder than the condensing surface temperature: an evaporative refrigeration effect. Such evaporative refrigeration effect may lead to new air-conditioning and refrigeration technologies, which is especially important considering current air-conditioning and refrigeration systems rely heavily on the Joule-Thomson effect, which require compressors and employ refrigerants that are harmful to the environment. These predictions, however, are based on approximate interfacial conditions that neglect the transport within the Knudsen layer, and hence call for further validation \new{by experiments or more accurate simulations}. \new{In fact,}  most experimental results on evaporation so far showed that the vapor side temperature is higher than that of the liquid surface. \new{Although a} few existing experiments \new{indeed reported} the vapor side temperature \new{is} lower \new{than the liquid interface temperature, the reported values are} within less than $1^\circ\text{C}$ in the vapor side compared to the liquid interface~\cite{Zhu-2013-Interfacial, Zhu-2009, Gatapova-2024-PhysFL}.

\new{Since experiments remain an unsolved challenge, we set to further evaluate Chen's approximations~\cite{GC-2022-flux-bc, GC-2024-inversion, GC-2025, Chen-2023-parallel} by solving the BTE. Unlike previous BTE-based studies that neglect the liquid regions,}
in this work, \new{we study} the \new{classical} evaporation and condensation problem between two parallel plates \new{including the liquid regions by coupling} the \new{BTE} description for the vapor phase to the continuum treatment for the liquid films.  Interfacial molecular absorption, reflection, and evaporation are included via the boundary conditions coupling the liquid and the vapor regions. Energy balance is applied at the interface to self-consistently determine the interfacial temperatures of the liquid films and the vapor.  Our \new{simulation results showed that the semi-continuum approach established by Chen works well in the small Knudsen number ($Kn$) regime. When $Kn$ is larger, $Kn \sim 1$, small differences exist in numerical values between the semi-continuum approach and the BTE solutions, although the trends of both approaches are consistent. Our work confirms the evaporative refrigeration effect.  The degree of cooling depends on several key parameters, including the Knudsen number, the liquid film thickness, the vapor-gap thickness, and the accommodation coefficient. Our prediction should stimulate further experiments to validate this evaporative refrigeration effect.}

\section{\new{Mathematical Formulation}}%Method}
\begin{figure}[tbhp]
\centering
\includegraphics[width=\linewidth]{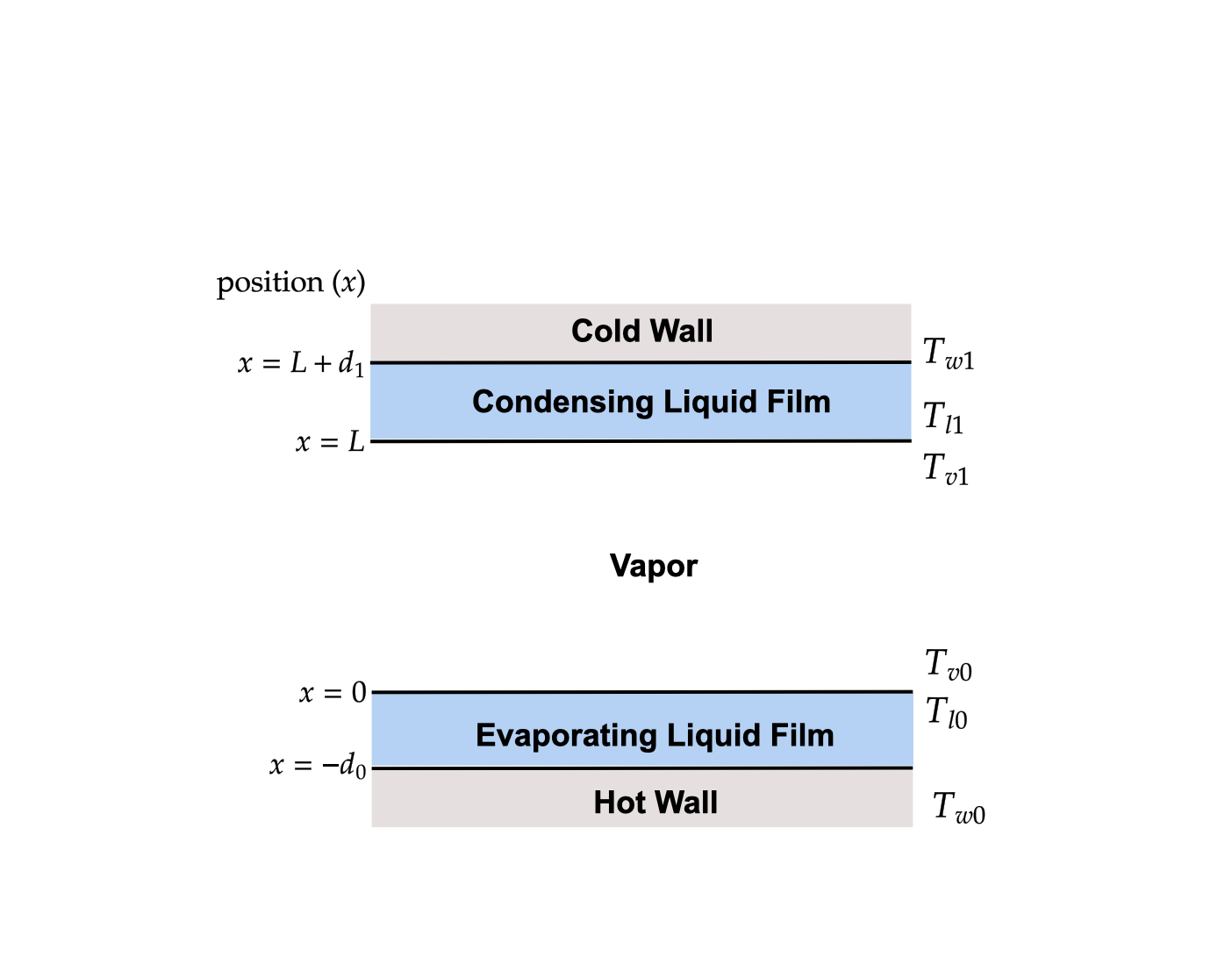}
\caption{\new{Schematic illustrating the simulated parallel plate configuration, including the liquid and vapor regions.}}
\label{fig: experiment_setup}
\end{figure}

\new{The details of equations and interfacial conditions are presented in this section. Figure~\ref{fig: experiment_setup} shows the simulated region and coordinate.} The model for the evaporation-condensation phenomenon in this setting is composed of three components: the temperature variation in (hot) liquid film is modeled by \new{heat conduction}, the vapor phase is modeled by the nonlinear \new{BTE}, and the temperature variation in (cold) liquid film is once again modeled by the diffusion equation. At the interfaces between liquid film and the vapor, energy conservation is imposed. $T_{w_0}$ and $ T_{w_1}$ denote the wall temperature at hot and cold wall respectively, and $T_{l_0} \new{(T_{v_0})}$ and $T_{l_1} \new{(T_{v_1})}$ denote the corresponding liquid \new{(vapor)} temperatures \new{at the hot and cold interfaces, respectively.}

The temperature variation within the liquid film is modeled by heat conduction, \new{while the convection effect can be neglected}, which has been justified~\cite{VC-2013-evaporation-potential-barrier, Chen-2023-parallel}, and the solution linearly interpolates the wall and the liquid interface temperature. Accordingly, the vapor heat flux at the evaporating and condensing are, respectively
\begin{align} \label{eqn: heat flux-liquid x0}
q_{\new{v}_0} = k_w \frac{T_{w_0}-T_{l_0}}{d_0} - \dot{m}\Gamma\,, \ 
q_{\new{v}_1} = k_w \frac{T_{l_1}-T_{w_1}}{d_1} - \dot{m} \Gamma\,,
\end{align}
where $k_w=0.6\mathrm{W}\cdot\mathrm{m}^{-1}\cdot\mathrm{K}^{-1}$ is heat conductivity of the liquid, $d_0$ and $d_1$ are thicknesses of liquid layers. Since liquid undergoes phase changes from liquid to vapor, the heat flux also composes a component $\dot{m}\Gamma$ with $\Gamma=2.3 \mathrm{M J}\cdot\mathrm{kg}^{-1}$ denoting the latent heat for the phase change and $\dot{m}$ representing the mass flux. 

In the vapor region, we deploy the steady \new{Bhatnagar-Gross-Krook(BGK)} model \new{of the BTE} to describe the gas dynamics\new{~\cite{AM-1994-evaporation-numerics, Pao-oppo-sign-1971}}. Since our system enjoys planetary symmetry, the drift velocity is assumed only in the direction perpendicular to the wall, and the $1D$-space $3D$-velocity model reads
\begin{align}\label{eqn: 3D-BGK}
v_1 \partial_x f & = \frac{1}{\tau}\left(\mathcal{M}[f]-f\right)\,, \quad x\in [0,L]\,, v=(v_1, v_2, v_3)\in \R^3 \,, 
\end{align}
here $L$ represents the thickness of vapor region, $f = f(x,v_1, v_2, v_3)$ is the mass density of gas molecules\new{, the number density multiplied by the typical mass,} at location $x$ with velocity $v = (v_1, v_2, v_3)$ \new{in $3\mathrm{D}$ space}, $\tau$ is the relaxation time which is treated as a constant here~\cite{Pao-oppo-sign-1971, SO-1978-kinetic, AM-1994-evaporation-numerics, GC-2022-flux-bc} and $\mathcal{M}[f]$ represents the local equilibrium (displaced Maxwellian):
\begin{equation}\label{eqn: Mf-def}
\mathcal{M}[f] = \frac{\rho}{\left(2\pi k_B T/m_0\right)^{3/2}} e^{-\frac{(v_1-u)^2 + v_2^2 + v_3^2}{2 k_B T/m_0}}\,,
\end{equation}
where $k_B$ is the Boltzmann constant, $m_0$ is the molecular mass, $\rho$ and $u$ represent the local \new{mass} density and drift \new{(or average)} velocity respectively. 
\new{It should be noted that the inclusion of translational energy only implies treatment of monatomic gas, even though liquid and latent heat values, and saturation properties are taken to be close to those of water. Due to this limitation, the results are intended to capture qualitative trends, rather than exact.  The relaxation time $\tau$ is taken as constant, as a more exact treatment of the relaxation time would only introduxe numerical differences without changing the trend.  The comparison with semi-continuum treatment, however, remains exact, since both approaches assume a monatomic gas, as will be explained later.}

\new{
In the remainder of this section, the symmetry of the BTE is exploited to reduce it to two coupled systems, after which the appropriate boundary conditions corresponding to our problem are specified. The decomposition of energy flux is introduced for use in later discussions, followed by the coupling between the BTE and the liquid regions. Finally, the nondimensionalization employed in the simulations is presented.
}

\subsection{Reduction of Dimension}

Due to the assumption of symmetry as in \new{Eq.}~\eqref{eqn: Mf-def} that drift velocity $u$ is in the first direction \new{($x$-direction in Fig.~\ref{fig: experiment_setup})}, the original velocity space can be reduced to $1D$ as \new{shown} in Aoki et al.~\cite{ASY-1990-condensed-numerics}, by multiplying $1$ and $(v_2^2 + v_3^2)$ to both sides in \new{Eq.}~\eqref{eqn: 3D-BGK} then integrate over $v_2$ and $v_3$. \new{Defining two} new density functions:
\begin{align}
g(x, v_1) &:= \int_{\R^2} f\,dv_2 dv_3\,, \label{eq:def_g_v1}\\
h(x, v_1) &:= \int_{\R^2} (v_2^2 + v_3^2)f \,dv_2 dv_3\,, \label{eq:def_h_v1}
\end{align}
\new{the above operation leads to the} following \new{equations}:
\begin{align}
v_1 \partial_{x} g &= -\frac{1}{\new{\tau}}
\left(g- \frac{\rho}{\sqrt{2\pi\theta}}e^{-\frac{-(v_1-u)^2}{2\theta}} \right)\,, \label{eqn: non-dimension-BGK-g}\\
v_1 \partial_{x} h& = -\frac{1}{\new{\tau}}\left(h - \frac{2 \rho \theta}{\sqrt{2\pi \theta}}e^{-\frac{(v_1-u)^2}{2\theta}} \right)\,, \label{eqn: non-dimension-BGK-h}
\end{align}
The macroscopic quantities of density, average velocity and \new{$\theta:=k_B T/m_0$}, are defined by \new{conservation of mass, momentum and energy, more specifically~\cite{AM-1994-evaporation-numerics},}
\begin{align}
\rho(x) &:= \int_{\R} g\,dv_1\,,\label{eqn:rho_def}\\
u(x) &:= \frac{1}{\rho}\int_{\R} v_1 g \,dv_1\,, \label{eqn:u_def}\\
\theta(x) &:= \frac{1}{3\rho} \int_{\R} ((v_1-u)^2 g + h)\,dv_1 \,. \label{eqn:theta_def}
\end{align}

The \new{BTE} is \new{subjected to the} boundary conditions~\new{\cite{GC-2022-flux-bc}}, 
\begin{equation}\label{eq:BC_f_alpha}
\begin{array}{rl}
f(x=0, v_1>0) & = \alpha \phi_0 + (1-\alpha)f(x=0, -v_1)\,,  \\
f(x=1, v_1<0) & = \alpha \phi_1 + (1-\alpha)f(x=1, -v_1)\,.  
\end{array}
\end{equation} 
\new{where $\phi_0$ and $\phi_1$ are Maxwellian distributions corresponding to the local liquid interface temperatures $T_{l_0}$ for the hot side and $T_{l_1}$ for the cold side, respectively, and} $\alpha$ \new{is} the accommodation coefficient. \new{The first term in Eq.~\eqref{eq:BC_f_alpha} represents vapor molecules leaving the liquid interface, and the second term represents the fraction of reflected molecules at the interface, assuming} specular reflection. \new{The specular assumption may lead to numerical difference from diffuse interface result, but the trend should be consistent. Most previous studies used this approximation, including the quasi-continuous treatment of Chen~\cite{GC-2022-flux-bc}.}

The boundary conditions for $g$ and $h$ are accordingly determined by their definitions \new{Eqs.}~\eqref{eq:def_g_v1} \new{and}~\eqref{eq:def_h_v1}, together with these we arrive at the final \new{equations to solve:}
\begin{equation}
\left\{\begin{array}{rl}
v_1 \partial_x g &= -\frac{1}{\new{\tau}}\left(g - \frac{\rho}{\sqrt{2\pi\theta}}e^{-\frac{-(v_1-u)^2}{2 \theta}} \right) \,, \label{eqn: final-BGK-g}\\
g(x=0, v_1>0) & = \alpha \phi_0^g + (1-\alpha)g(x=0, -v_1)\,,  \\
g(x=1, v_1<0) & = \alpha \phi_1^g + (1-\alpha)g(x=1, -v_1)\,, 
\end{array}\right.
\end{equation}
\begin{equation}
\left\{\begin{array}{rl}
v_1 \partial_x h &= -\frac{1}{\new{\tau}}\left(h - \frac{2\rho\theta}{\sqrt{2\pi \theta}}e^{-\frac{(v_1-u)^2}{2 \theta}} \right)\,, \label{eqn: final-BGK-h} \\
h(x=0, v_1>0) & = \alpha \phi_0^h + (1-\alpha)h(x=0, -v_1)\,,\\  %\label{eqn: steady-BGK-h-BC0}\\
h(x=1, v_1<0) & = \alpha \phi_1^h + (1-\alpha)h(x=1, -v_1)\,. %\label{eqn: steady-BGK-h-BC1}
\end{array}\right.
\end{equation}
When $\alpha=1$ the boundary condition\new{s are} reduced to Dirichlet type~\cite{AM-1994-evaporation-numerics}, whereas $\alpha=0$ corresponds to only \new{complete} specular reflection. 

As a consequence of the model above, the mass flux and energy flux in the vapor region are computed by:
\begin{align}
\dot{m}(x) &= \int_{\R^3} v_1 f\,dv = \int_{\R} v_1 g\,dv_1\,, \label{eqn: mass flux-v}\\
j_e(x) &= \int_{\R^3} v_1 \frac{|v|^2}{2} f\,dv = \int_{\R} \frac{v_1^3}{2} g\,dv_1 + \int_{\R} \frac{v_1}{2} h\,dv_1\,. \label{eqn: heat flux-v}
\end{align}

\subsection{Decomposition of Energy Flux \label{sec:decompose_energy_flux}} 
Total energy flux can be decomposed into heat flux, enthalpy flux, and kinetic energy flux. This can be seen from the steady state \new{BTE}~\eqref{eqn: 3D-BGK}, or equivalently~\eqref{eqn: non-dimension-BGK-g} and~\eqref{eqn: non-dimension-BGK-h}.

More specifically, \new{BTE under constant relaxation time approximation} preserves energy \new{conservation}, meaning:
\begin{equation}
\partial_x j_e = \partial_x \int_{\R^3} v_1 \frac{|v|^2}{2} f\,dv = 0\,. \label{eqn:energy_conserve}
\end{equation}
Expanding \new{the molecular energy} with respect to the molecular average velocity $u$, \new{i.e., we rewrite $v_1$ as $(v_1-u) + u$, to obtain}:
\begin{equation}
\begin{aligned}
j_e=& \frac{1}{2}\int_{\R^3} (v_1-u) ((v_1-u)^2 + v_2^2 + v_3^2) f\,dv \\
&+ \frac{1}{2}\int_{\R^3} (2(v_1-u) + u) u^2 f\,dv\\
&+ \frac{1}{2}\int_{\R^3} (v_1-u) (2u(v_1-u) + u^2 ) f\,dv \\
&+ \frac{1}{2}\int_{\R^3} u((v_1-u)^2 + v_2^2 + v_3^2) f\,dv\,.
\end{aligned}
\label{eqn:energy_conserve_trans}
\end{equation}
The first term is the microscopic expression of heat flux~\cite{vincenti1965introduction}:
\begin{align}\label{eqn:def_heat_flux}
q(x) = \frac{1}{2}\int_{\R^3} (v_1-u_1)((v_1-u_1)^2 + v_2^2 + v_3^2) f \,dv\,,
\end{align}
and the second term
\begin{equation}\label{eq:macro_kinetic_flux}
\frac{1}{2}\int_{\R^3} (2(v_1-u) + u) u^2 f\,dv= \frac{1}{2}\rho u^3\,,
\end{equation}
is understood as kinetic energy flux. To interpret the last two terms of~\eqref{eqn:energy_conserve_trans}, we notice:
\begin{equation}
\int_{\R^3} (v_1-u) (2u(v_1-u) + u^2 ) f\,dv = 2 u(\rho \theta) = 2 u p\,,
\end{equation}
where $p$ is the local pressure according to the ideal gas law, making this term represent the pressure work flux, and
\begin{equation}
\int_{\R^3} u((v_1-u)^2 + v_2^2 + v_3^2) f\,dv = 2u \rho e\,,
\end{equation}
represents the internal energy flux with $e$ denoting the internal energy. Considering enthalpy $\mathsf{h}=p/\rho + e$~\cite{vincenti1965introduction}, we can understand the last two terms together as the enthalpy flux $\rho uh$. Therefore \new{the energy conservation equation~\eqref{eqn:energy_conserve} can be understood as} 
\begin{equation}
\partial_x j_e=
\partial_x \left(q + \rho u \mathsf{h} + \frac{1}{2} \rho u^3\right) = 0\,.
\end{equation}

\subsection{Coupling of \new{BTE to Liquid Regions}}
At the interface of liquid film and water vapor, the diffusion equation for the liquid is coupled with the \new{BTE} for the vapor through energy conservation.

\subsubsection{Liquid to vapor}
Liquid temperature at the interface provides boundary condition for the \new{BTE}. More specifically, $\phi_{0,1}$ are assumed to be the local equilibrium under saturation temperature $T_{l_0}, T_{l_1}$ respectively, or in $\theta_{l_0}=k_B T_{l_0}/m_0, \theta_{l_1}=k_B T_{l_1}/m_0$,
\begin{equation}\phi_{0} = \frac{\rho_0}{(2\pi\theta_{l_0})^{3/2}}e^{-\frac{v_1^2 + v_2^2 + v_3^2}{2\theta_{l_0}}}\,,\ 
\phi_{1} = \frac{\rho_1}{(2\pi\theta_{l_1})^{3/2}}e^{-\frac{v_1^2 + v_2^2 + v_3^2}{2\theta_{l_1}}}\,, \end{equation}
and hence
\begin{equation}
\phi_i^g = \frac{\rho_{i}}{\sqrt{2\pi\theta_{l_i}}}e^{-\frac{v_1^2}{2\theta_{l_i}}}\,, \ 
\phi_i^h = \frac{2\rho_{i}\theta_{l_i}}{\sqrt{2\pi\theta_{l_i}}}e^{-\frac{v_1^2}{2\theta_{l_i}}}\,, 
\quad i=0,1 \,,
\end{equation}
where the interface densities $\rho_{0,1}$ are the dimensionless densities $\rho_{0,1} = \rho(T_{l_0,l_1})/\rho_r$, with $\rho(T)$ being determined by the Clausius-Clapeyron equation for saturated gas: 
\begin{equation}%\label{eqn: CC def}
\rho = \dfrac{\rho_r T_r}{T} \exp{\left(-\frac{\Gamma}{R}\left(\frac{1}{T}-\frac{1}{T_r}\right)\right)}\,,
\end{equation}
where $R$ is the ideal gas constant, $\Gamma$ is the latent heat, and $(\rho_r, T_r)$ is the density and temperature at some reference point, and take on fixed values for specific material.

For example, for water close to room temperature, setting $T_r = 273\mathrm{K}$, this nonlinear relation is approximately~\cite{Chen-2023-parallel}:
\begin{equation}\label{eqn: nonlinear CC}
\rho = 10^{-3}\left[5.018 + 0.3232\Delta T + 8.1847\times10^{-3} (\Delta T)^2 + 3.1243\times 10^{-4}(\Delta T)^3\right]\,,
\end{equation}
where $\Delta T = T-273$ represents the temperature difference between absolute temperature and $T_r$.

\subsubsection{Energy balance}
The vapor \new{BTE} provides boundary condition for liquid as well. This condition comes in the form of energy conservation.

Recall~\eqref{eqn: heat flux-liquid x0} derived from the heat equation for the liquid, and~\eqref{eqn: heat flux-v} derived from \new{BTE} model for vapor, we equip the two sources of energy flux to obtain:
\begin{equation}\label{eqn: flux-conserve}
\left\{\begin{array}{ll}
k_w \dfrac{T_{w_0}-T_{l_0}}{d_0} - \dot{m}(x=0)\Gamma = j_e(x=0) \,,\\
k_w \dfrac{T_{l_1}-T_{w_1}}{d_1} - \dot{m}(x=L)\Gamma = j_e(x=L)\,.
\end{array}\right.\end{equation}
Noting that $j_e$ and $\dot{m}$ are solutions to \new{Eqs.}~\eqref{eqn: final-BGK-g} \new{and}~\eqref{eqn: final-BGK-h}, hence \new{Eq.}~\eqref{eqn: flux-conserve} presents a highly nonlinear convoluted system. We solve it in an iterated scheme. At the $n$-th iteration, the current saturation temperature $(T^n_{l_0}\,,T^n_{l_1})$ are utilized to solve~\eqref{eqn: final-BGK-g}~\eqref{eqn: final-BGK-h} under the boundary conditions. The outputs of the solutions, denoted by $g^n$ and $h^n$, are used to compute $\dot{m}^n, j_e^n$ in \new{Eqs.}~\eqref{eqn: mass flux-v} \new{and}~\eqref{eqn: heat flux-v}, then to update the \new{new values} as in \new{Eq.}~\eqref{eqn: fixed pt sol}.
\begin{figure*}[htbp]
\begin{equation}\label{eqn: fixed pt sol}
\left\{\begin{array}{ll}
T_{l_0}^{n+1} = T_{w_0} - \frac{d_0}{k_w} \Gamma \int_{\R} v_1 g^n(x=0)\,dv_1 
- \frac{d_0}{k_w} \frac{1}{2} \int_{\R} v_1^3 g^n(x=0)\,dv_1 
- \frac{d_0}{k_w} \frac{1}{2} \int_{\R} v_1 h^n (x=0)\,dv_1\,,\\
T_{l_1}^{n+1} = T_{w_1} + \frac{d_1}{k_w} \Gamma \int_{\R} v_1 g^n(x=L)\,dv_1 + \frac{d_1}{k_w} \frac{1}{2} \int_{\R} v_1^3 g^n(x=L)\,dv_1
+ \frac{d_1}{k_w} \frac{1}{2} \int_{\R} v_1 h^n (x=L)\,dv_1\,.
\end{array}\right.\end{equation}
\end{figure*}

\new{The convergence criterion of this iteration scheme is the relative error between $T_{l_0}^{n+1}$ and $T_{l_0}^{n}$, and between $T_{l_1}^{n+1}$ and $T_{l_1}^{n}$ satisfies $\frac{T_{li}^{n+1}-T_{li}^n}{T_{li}^n}<10^{-6}$. Additional criteria have also been tested, including longer simulation time and more refined time steps, and the fixed-point solution was found to be robust. Variations in runtime and time-step size have no significant change in the results.}

\subsection{Semi-continuum solution} \label{subsec:semi_sol}
The semi-continuum solution is based on the asymptotic computations in the Supplementary materials, introduced by Chen~\cite{GC-2022-flux-bc}. \new{To be consistent with the monatomic approximation implied in the Maxwellian velocity distribution, we use standard results from the kinetic theory for the specific heat and thermal conductivity for monatomic gas~\cite{GC-book-2005}:} $c_p = \frac{5}{2}\frac{k_B}{m_0} = 1153 \,\mathrm{J}\cdot\mathrm{K}^{-1}\cdot\mathrm{kg}^{-1}, k_v = \frac{5}{2}\tau \left(\frac{k_B}{m_0}\right)^2\rho T 
= 0.0036\,\mathrm{W}\cdot\mathrm{K}^{-1} \cdot\mathrm{m}^{-1},$
\new{where} $\tau=0.5\times 10^{-9}\,\mathrm{s}$ \new{is taken. Since the vapor pressure in the vapor region is constant in the semi-continuum approximation, it is set that} $\rho T=13.6\,\mathrm{kg}\cdot\mathrm{m}^{-3}\cdot\mathrm{K}$.

\subsection{\new{Nondimensionalization}}\label{subsec:nondim_numerics}
\new{We solve equations~\eqref{eqn: final-BGK-g} and~\eqref{eqn: final-BGK-h} numerically using a finite difference scheme in MATLAB. For the convenience of numerical computation and subsequent discussions, the following normalized variables are introduced:}
$\hat{x} :=x/L\,, \hat{v}_i:=v_i/u_{\new{c}}\,,  
\hat{\theta}:= \frac{\new{\theta}}{u_{\new{c}}^2}\,,$
where $u_{\new{c}}$ \new{is a characteristic velocity of the order of $\sqrt{2k_B T/m_0}$, which is taken as $u_{\new{c}}=200\mathrm{m}\cdot\mathrm{s}^{-1}$.} \new{Using} these new variables, $\hat{f}(\hat{x}, \hat{v}_i) = f(L\hat{x}, u_{\new{c}}\hat{v}_i) = f(x,v)$ satisfies the dimensionless \new{BTE}:
\begin{equation} \label{eqn: non-dimension-BGK-f}
\hat{v}_1\partial_{\hat{x}}\hat{f}(\hat{x}, \hat{v}_i) = -\dfrac{1}{\tau u_{\new{c}}/L} \left(\hat{f}(\hat{x}, \hat{v}_i) - \mathcal{M}[\hat{f}]\right)\,,
\end{equation}
for $0<\hat{x}<1, \hat{v} \in \R^3$ \new{($\hat{v}$ in three-dimensional space)} and $Kn:=\tau u_c/L$ denoting the Knudsen number that represents the ratio between molecular mean free path and the characteristic length of the vapor region.

\section{Results}

\begin{figure*}
\begin{subfigure}[b]{0.32\textwidth}
    \centering
    \includegraphics[width=\textwidth]{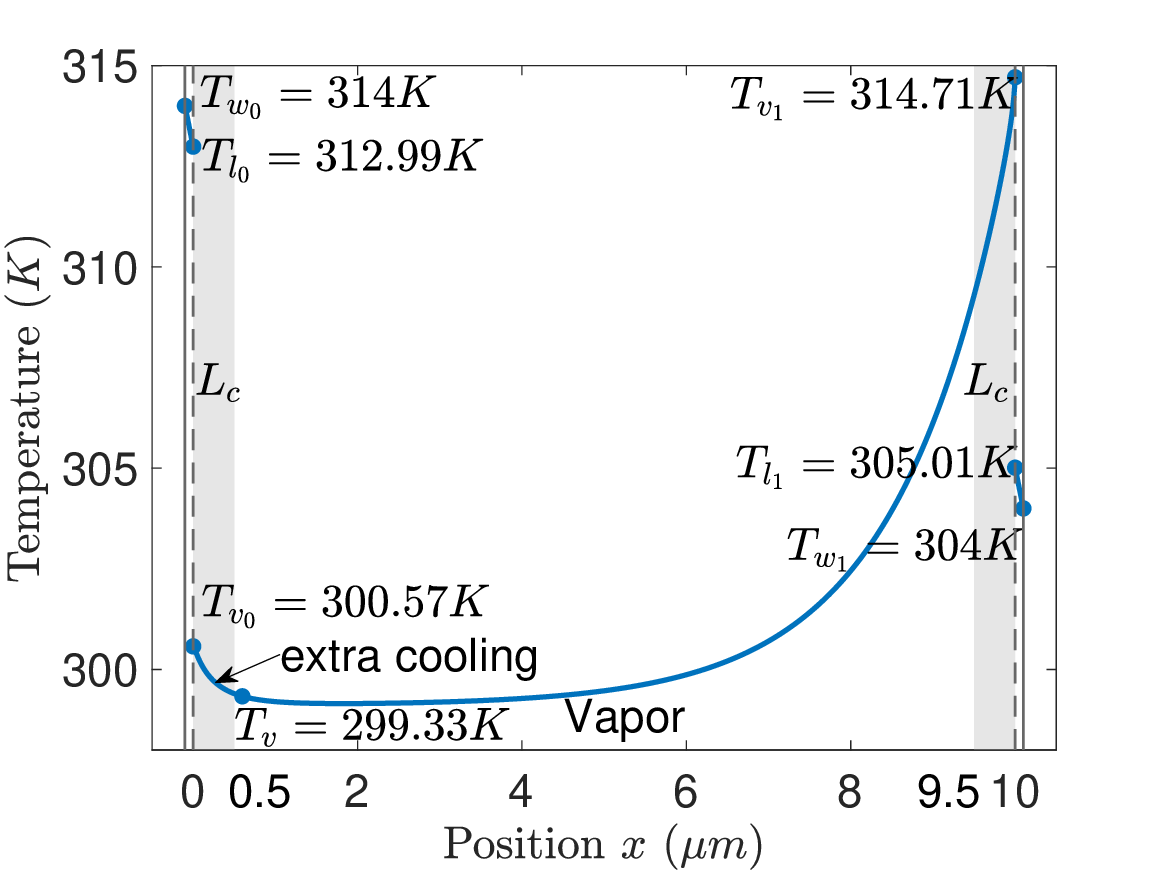}
    \caption{ }
    \label{fig:Temp_dist_example}
\end{subfigure}
     \hfill
\begin{subfigure}[b]{0.32\textwidth}
    \centering
    \includegraphics[width=\textwidth]{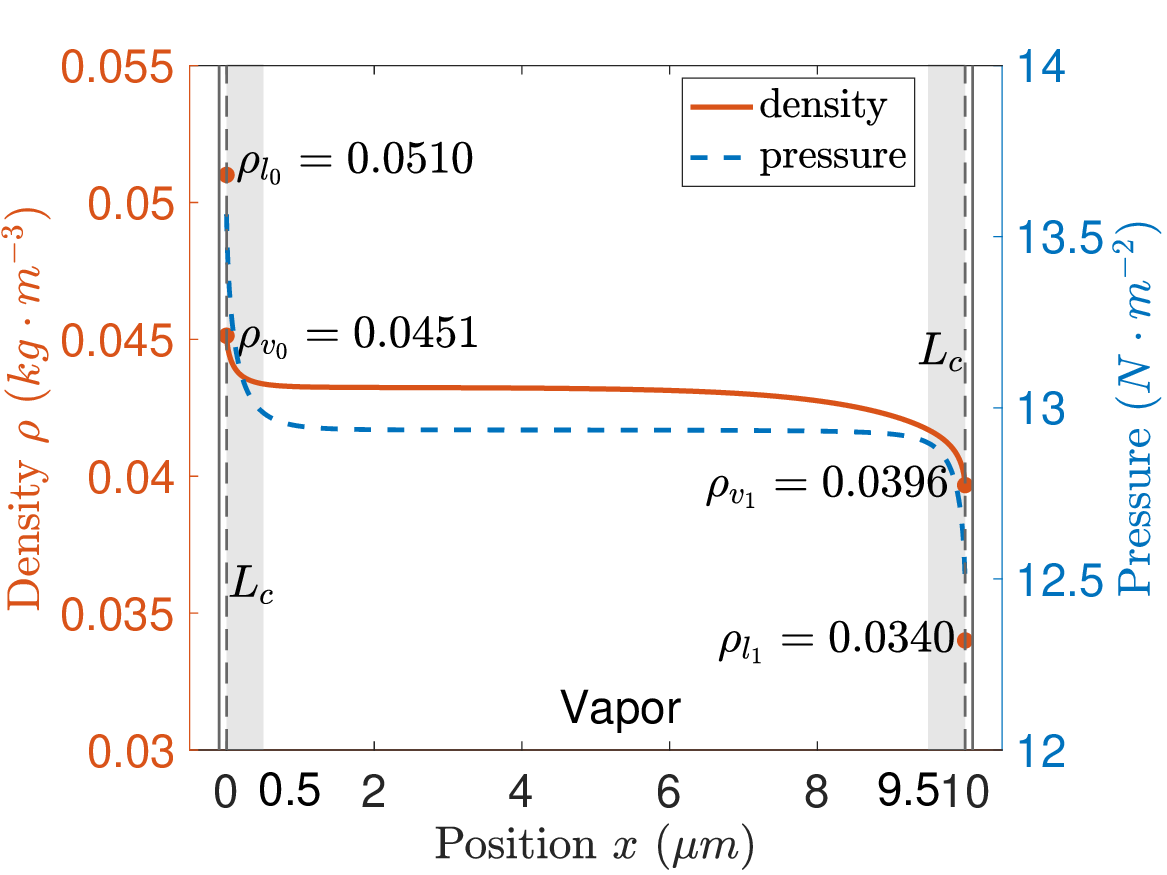}
    \caption{ }
    \label{fig:rho_dist_example}
\end{subfigure}
    \hfill
\begin{subfigure}[b]{0.32\textwidth}
         \centering
         \includegraphics[width=\textwidth]{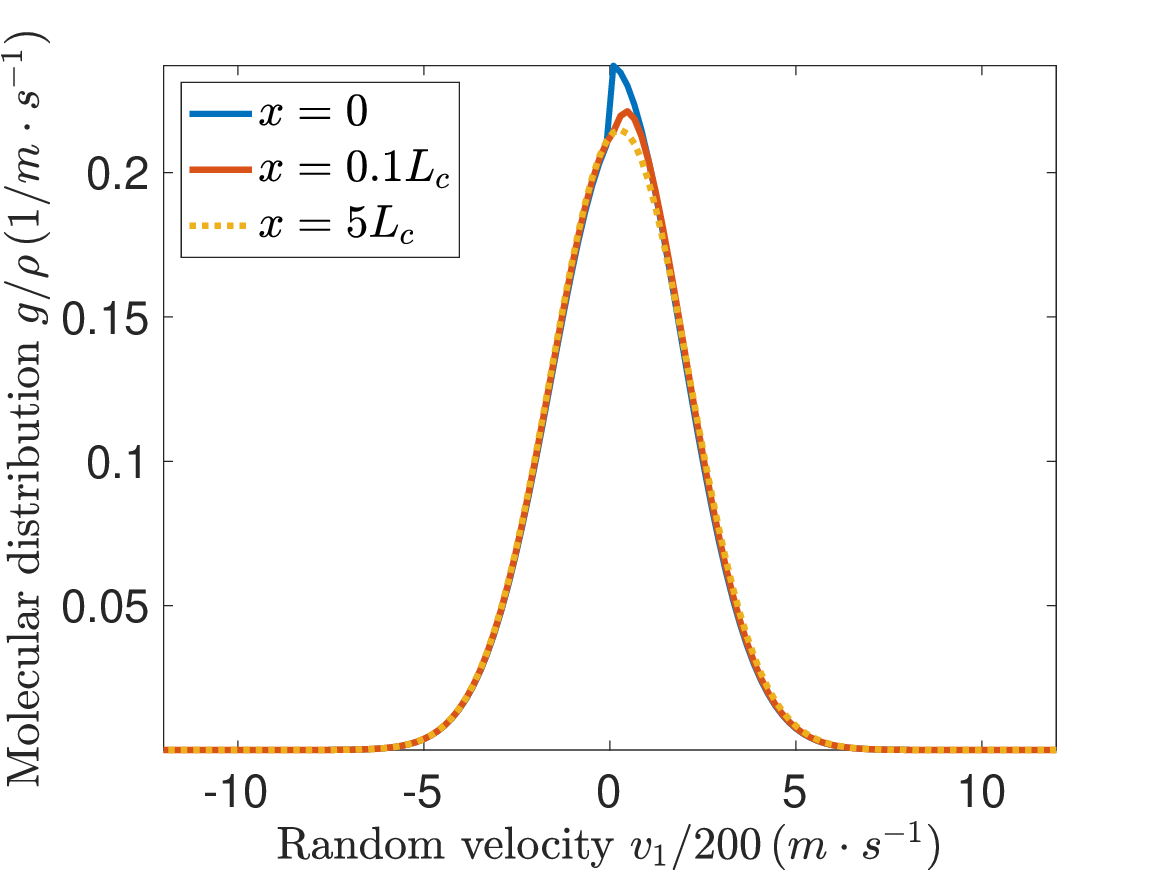}
    \caption{ }
    \label{fig:dist_example}
\end{subfigure}
    \\
\begin{subfigure}[b]{0.32\textwidth}
         \centering
         \includegraphics[width=\textwidth]{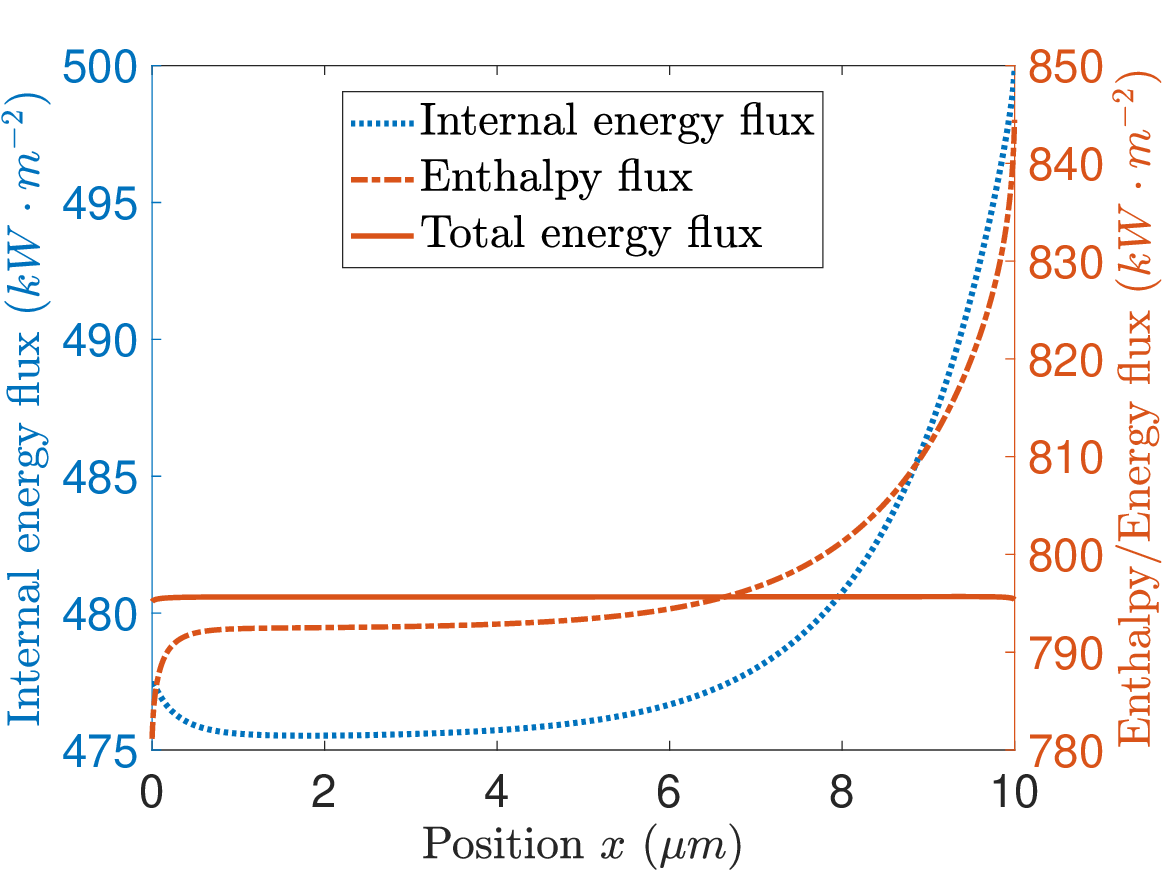}
    \caption{ }
    \label{fig:energy_flux_example}
\end{subfigure}
    \hfill
\begin{subfigure}[b]{0.32\textwidth}
         \centering
         \includegraphics[width=\textwidth]{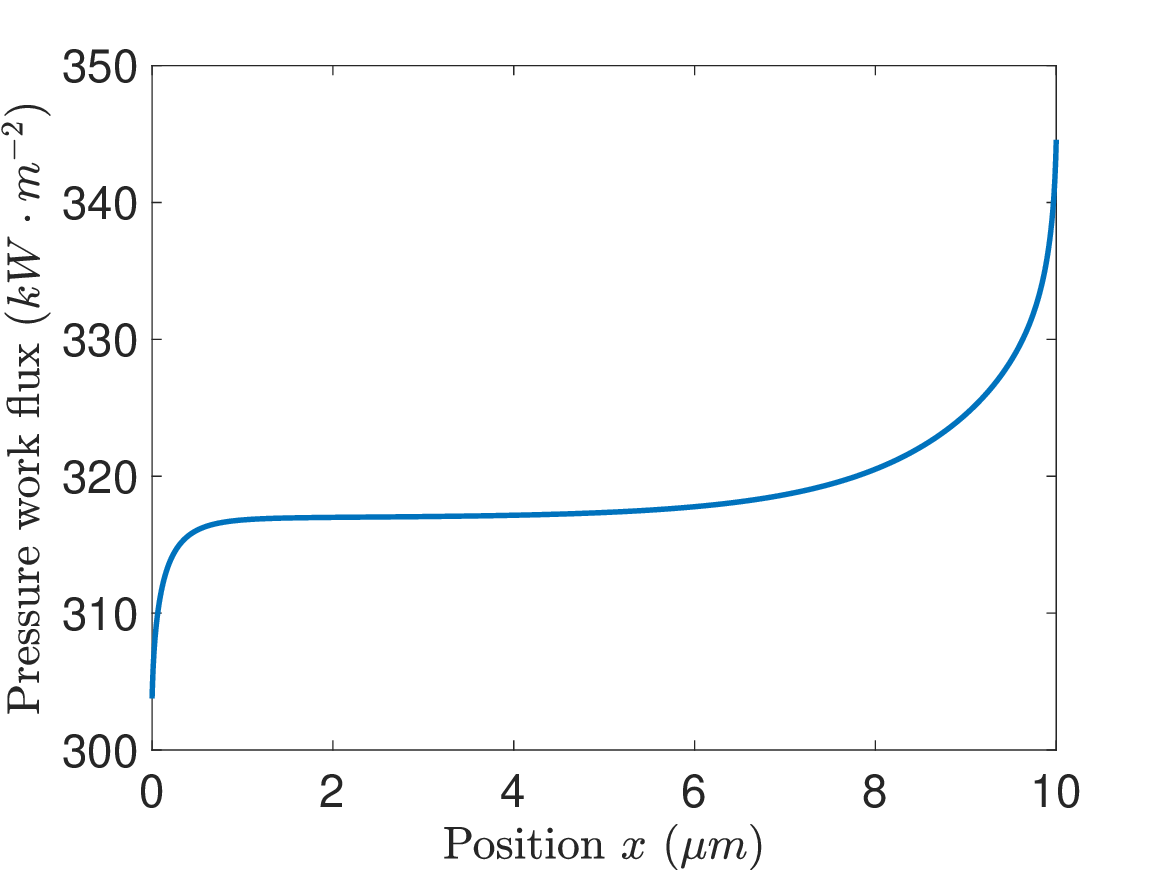}
    \caption{ }
    \label{fig:pressure_work_example}
\end{subfigure}
    \hfill
\begin{subfigure}[b]{0.32\textwidth}
         \centering
         \includegraphics[width=\textwidth]{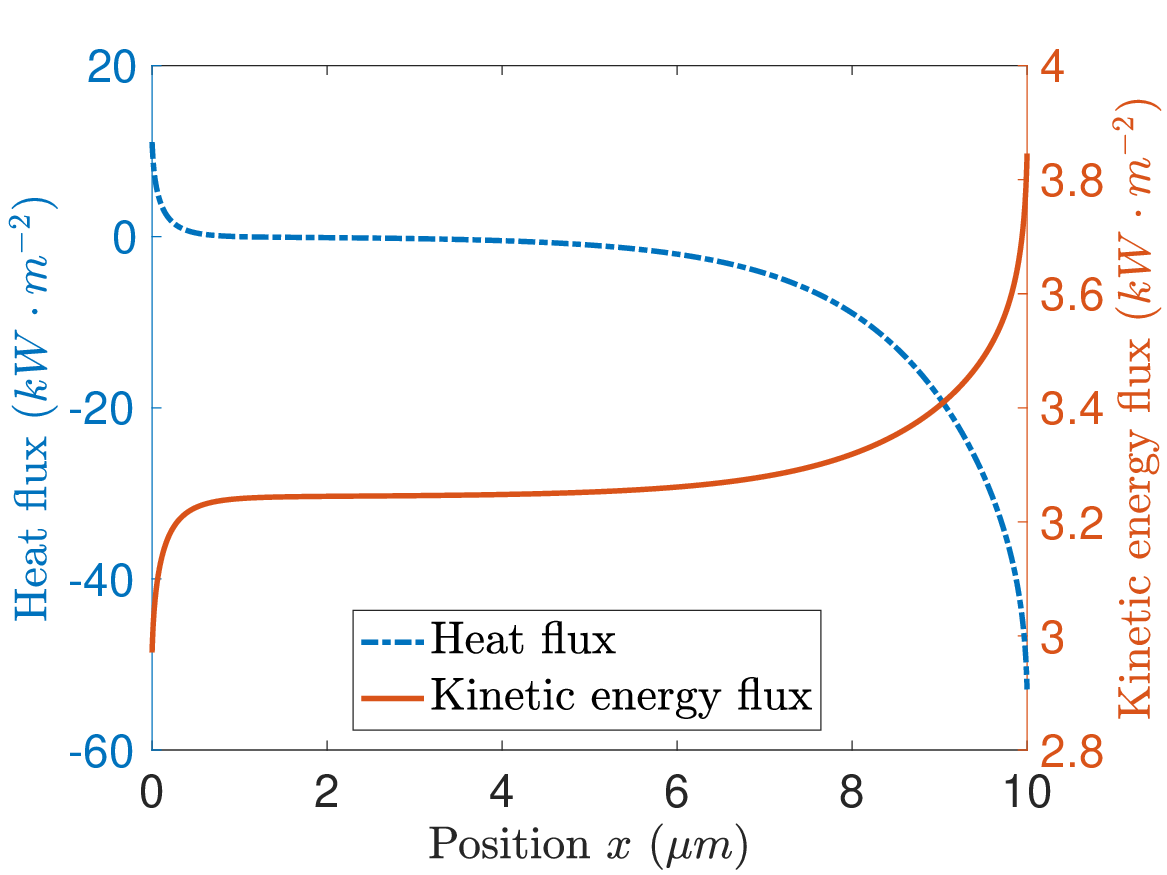}
    \caption{ }
    \label{fig:heat_flux_example}
\end{subfigure}
    \caption{\new{Refrigeration effect and mechanisms. 
    (a) Temperature distribution showing the inversion in the vapor region and discontinuities at the interfaces. The lowest temperature is about $5\mathrm{K}$ lower than the cold wall temperature, i.e., a refrigeration effect. 
    (b) Distributions of density $\rho$ and pressure $\rho T$.
    (c) Molecular distribution function at the hot liquid-vapor interface $x=0$, in the Knudsen layer $x=0.1L_c$ and outside the Knudsen layer at $x=5L_c$. The distribution at the interface is highly asymmetric between the molecules leaving and coming towards the interface, which leads to the cooling at the interface.
    (d Distributions of total energy flux, enthalpy flux, and internal energy flux in the vapor region.
    (e) The distribution of pressure work flux in the vapor region. The additional cooling beyond the interface is due to the pressure work increase.
    (f) Distributions of heat flux and kinetic energy flux in the vapor region.
    All for $Kn=0.01, d=0.1\mu\mathrm{m},\alpha=1, T_{w_0}=314\mathrm{K}, T_{w_1}=304\mathrm{K}$. }}
    \label{fig: refrigeration show example}
\end{figure*}

Figure~\ref{fig:Temp_dist_example} \new{shows} one example \new{of the numerical solutions. In this example, the Knudsen number $Kn=0.01$ means that the vapor region is much larger than the mean free path, i.e., the transport is close to the continuum. The thickness of the liquid film on both sides is taken as the same $d_0 = d_1 = 0.1\mu\mathrm{m}$. The accommodation coefficient $\alpha$ is taken to be one, i.e., fully accommodating liquid surfaces without reflection.}
The wall temperature on the evaporation side is \new{set to} $T_{w_0}=314\mathrm{K}$ and the condensation side is \new{set to} $T_{w_1}=304\mathrm{K}$. \new{First,} we notice that the predicted temperature distribution in the vapor region is inverted: hotter at the condensation side than the evaporation side, consistent with the predictions made by Pao~\cite{Pao-oppo-sign-1971} and confirmed by others. \new{Second, large} temperature discontinuities exist at \new{both} interface\new{s}. \new{The vapor temperature at the hot side is lower than the evaporating liquid surface temperature, while the vapor at the cold side is higher than the condensing liquid surface temperature. Importantly,} the vapor temperature in most of the gap region is below the condensing wall temperature, as low as $299.15 \mathrm{K}, 5\mathrm{K}$ lower than the cold side temperature. This is clearly a refrigeration effect.  
% Fig.~\ref{fig:rho_dist_example} also plots the interfacial saturation vapor density, which differs from the vapor side density at the interface, i.e., vapor density discontinuities exist.

\new{Since there is no sharp transition between the Knudsen layer and the bulk region, here we will loosely define the Knudsen layer as $L_c\sim5\tau u_c$, i.e., five mean free paths. With the parameters chosen in section~\ref{subsec:semi_sol} and~\ref{subsec:nondim_numerics}, $L_c=500n\meter$. Figure~\ref{fig:Temp_dist_example} marks down $L_c$.}

\new{Fig.~\ref{fig:rho_dist_example} presents the density and pressure profile in the vapor region. Also plotted in Fig.~\ref{fig:rho_dist_example} are the interfacial saturation vapor densities, which differ from the vapor side densities at both interfaces, i.e., vapor density discontinuities at both interfaces exist. Unlike temperature, density drops from the hot side to the cold side. The opposing temperature and density trend makes the product $\rho T$, that is, pressure approximately constant in the bulk vapor region, which validates the approximations made in the semi-continuum treatment. The opposing trends of temperature and density result from the compressibility of the vapor phase. For such a problem, clearly the constant density approximation is not valid. This is an important insight for phase change heat transfer.}

Chen~\cite{GC-2024-inversion} discussed the origin of the refrigeration effect as arising from the difference in the carrier distribution function at the interfacial region and the continuum region.  At the interfacial region, vapor enthalpy in the forward direction is $2RT$, where $R$ is the gas constant. While outside the Knudsen layer, the vapor enthalpy is $2.5RT$, i.e., the constant pressure specific heat of a monatomic gas. The $2RT$ is also consistent with thermionic emission~\cite{Mahan-1994-AppPhys}, confirming the similarities between electron and molecular emission~\cite{GC-2022-flux-bc}. However, because Chen's treatment neglected the thickness of the Knudsen layer, he could not pinpoint exactly where cooling happens.

In Fig.~\ref{fig:dist_example}, the molecular distribution functions at the interface \new{$x=0, x=0.1 L_c$ and $x=5L_c$} are shown. \new{Compared with the distribution away from the interface, shown in dashed orange line, the distribution at the interface ($x=0$), indicated by the solid blue line, is much more asymmetric.} The forward half ($v_1>0$), representing the molecules leaving the surface, is clearly different from the incoming half ($v_1<0$). The forward enthalpy is $2RT$ while the distribution incoming to the interface is closer to the diffusion distribution outside the Knudsen layer. This asymmetry in the distribution function creates a large interfacial temperature drop, as is clear in Fig.~\ref{fig:Temp_dist_example}. This mismatch in the forward-going flux and the backward-going flux is the main mechanism for the drop of the vapor phase temperature below that of the liquid interface temperature, and it is the main contributor to the interfacial evaporative refrigeration effect.

Furthermore, Fig.~\ref{fig:Temp_dist_example} also shows that the vapor temperature drops slightly in the Knudsen layer, \new{as demonstrated by the temperature difference of about $1 \mathrm{K}$ between $T_{v_0}$ and $T_v$ at $x=L_c$}. To understand the origin of this additional temperature drop, we show in Fig.~\ref{fig:energy_flux_example} the total energy flux \new{(solid red line)}, the enthalpy flux \new{(dashed red line)}, and the internal energy flux \new{(dotted blue line)}. We note that the total energy flux is a constant, as expected. The total energy flux can be expressed as the sum of three parts: (1) the heat flux, (2) the enthalpy flux, and (3) the kinetic energy flux. The enthalpy flux can be further decomposed into the sum of the internal energy flux and the pressure work flux. Microscopic expressions for each of these fluxes are given in section~\ref{sec:decompose_energy_flux}. In the Knudsen layer near the evaporation side, we see that the enthalpy flux increases while the internal energy flux decreases away from the interface.  The difference of the two gives the pressure work flux, shown in Fig.~\ref{fig:pressure_work_example}, which increases away from the interface \new{since pressure $p$ is approximately constant and density $\rho$ decreases}.  The decreasing internal energy flux leads to additional lower temperature of the vapor phase, i.e., the additional cooling effect.  This effect bears similarities to the Joule-Thomson effect in a throttle valve.  However, the Joule-Thomson expansion is an isenthalpic process, while the expansion across the Knudsen layer in the evaporation side is not isenthalpic.  \textbf{In fact, the decreasing temperature also leads to decreasing heat flux away from the interface, i.e., more heat conducts into the Knudsen region than conducts out of the region, which partially diminishes the internal energy flux reduction and hence the additional cooling effect across the Knudsen layer.} \new{As shown in Fig.~\ref{fig:heat_flux_example}, the heat flux $q$ [Eq.~\eqref{eqn:def_heat_flux}] in the dashed blue line decreases at the interface $x=0$ near the hot wall, and is about a constant for the bulk vapor part, while across the Knudsen layer near the hot wall, the kinetic energy flux~\eqref{eq:macro_kinetic_flux} in solid red line increases.} However, under proper conditions, the additional cooling happens.

\begin{figure*}
\centering
\begin{subfigure}[b]{0.32\textwidth}
         \centering
         \includegraphics[width=\textwidth]{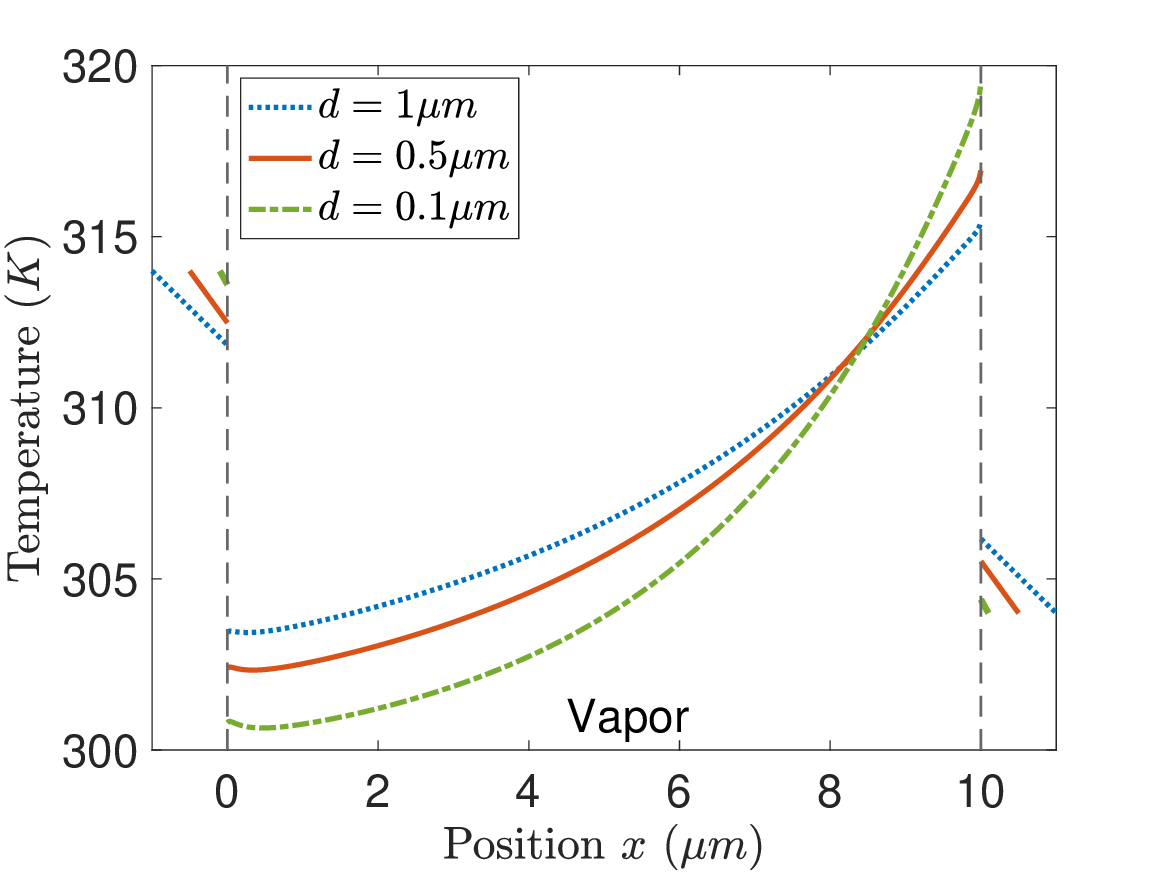}
    \caption{ }
    \label{fig:Temp_change_d}
    \end{subfigure}
    \hfill
    \begin{subfigure}[b]{0.32\textwidth}
         \centering
         \includegraphics[width=\textwidth]{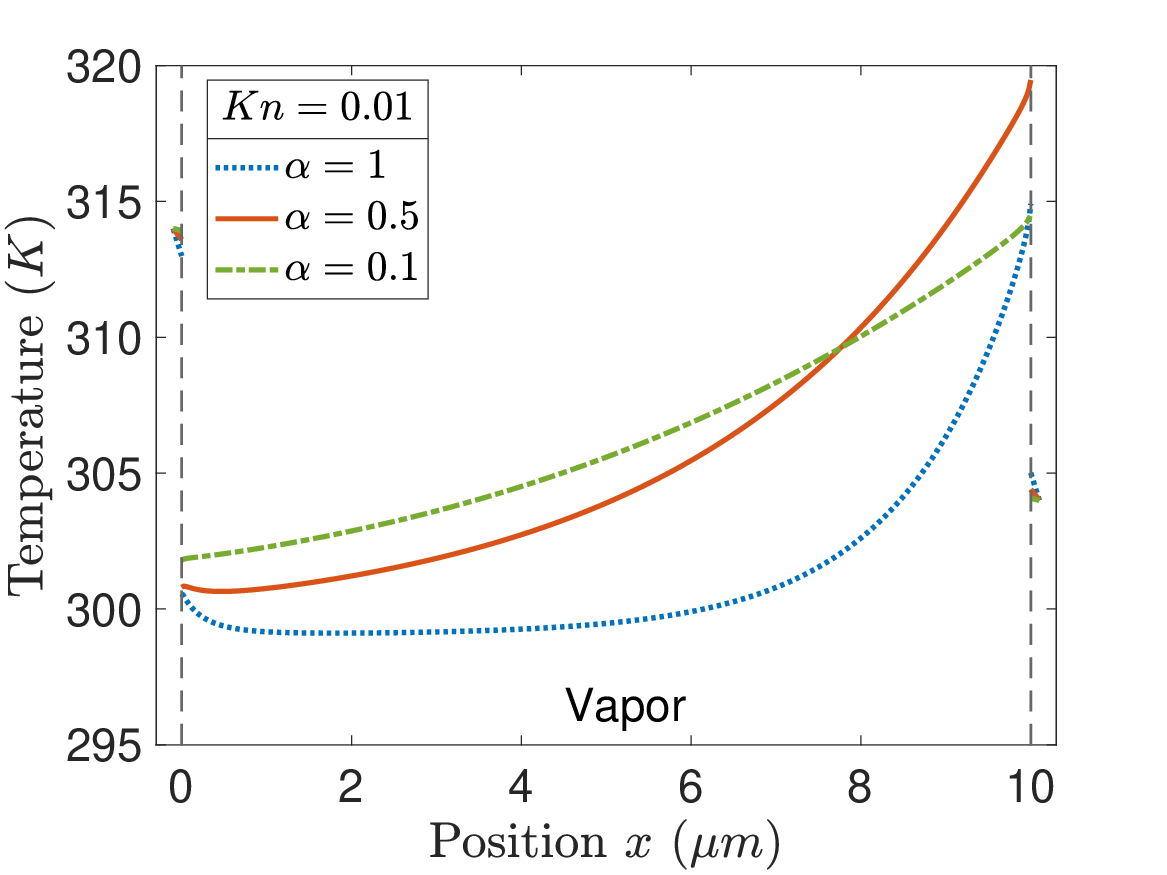}
    \caption{ }
    \label{fig:Temp_change_a_Kn001}
     \end{subfigure}
     \hfill
     \begin{subfigure}[b]{0.32\textwidth}
         \centering
         \includegraphics[width=\textwidth]{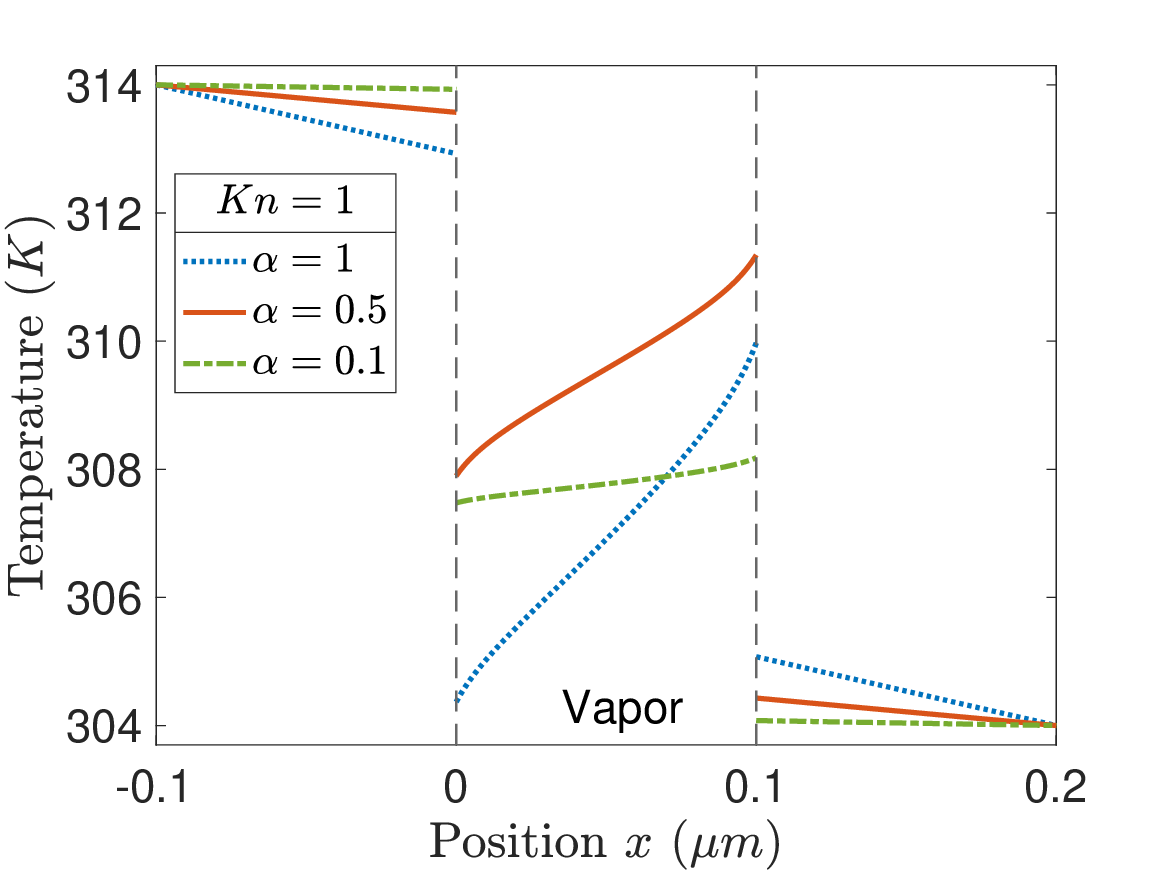}
    \caption{ }
    \label{fig:Temp_change_a_Kn1}
     \end{subfigure}
     \\
     \begin{subfigure}[b]{0.32\textwidth}
         \centering
         \includegraphics[width=\textwidth]{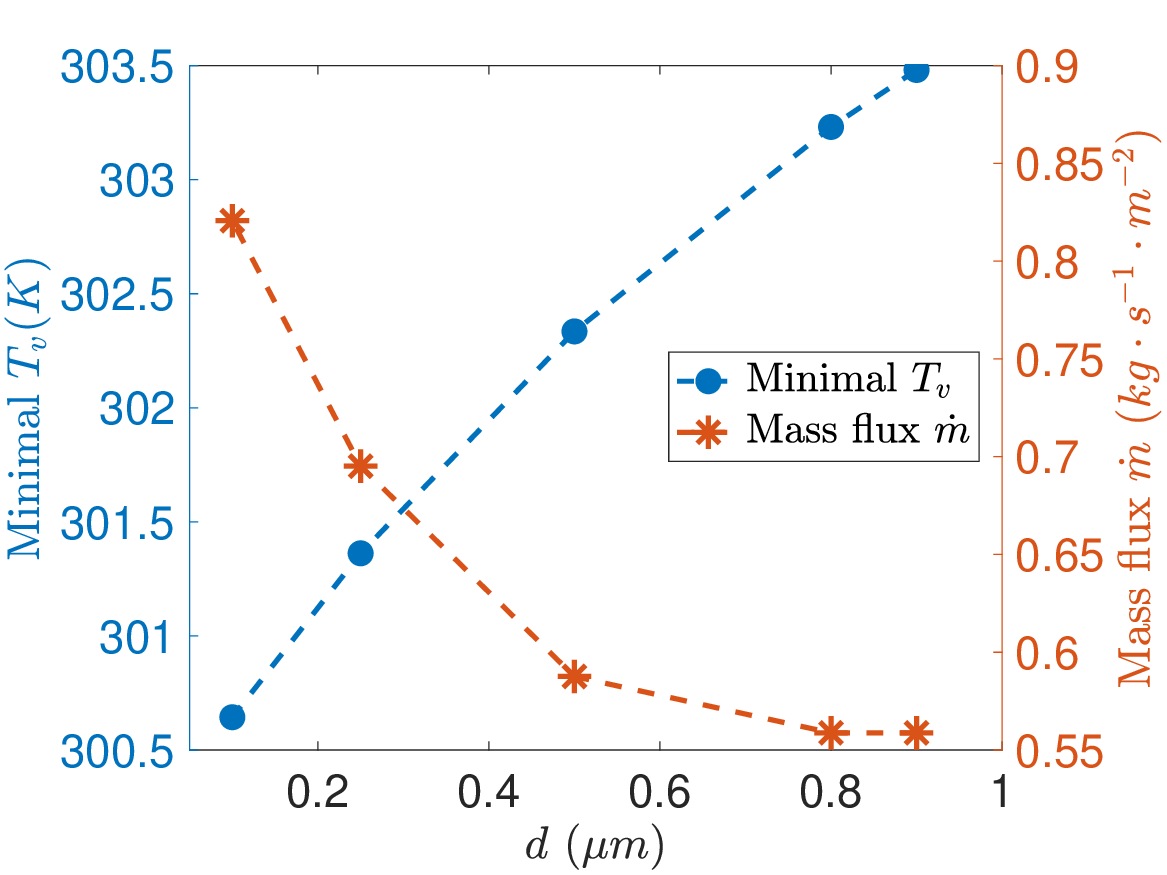}
    \caption{ }
    \label{fig:Tmin_massf_change_d}
    \end{subfigure}
    \hfill
     \begin{subfigure}[b]{0.32\textwidth}
         \centering
         \includegraphics[width=\textwidth]{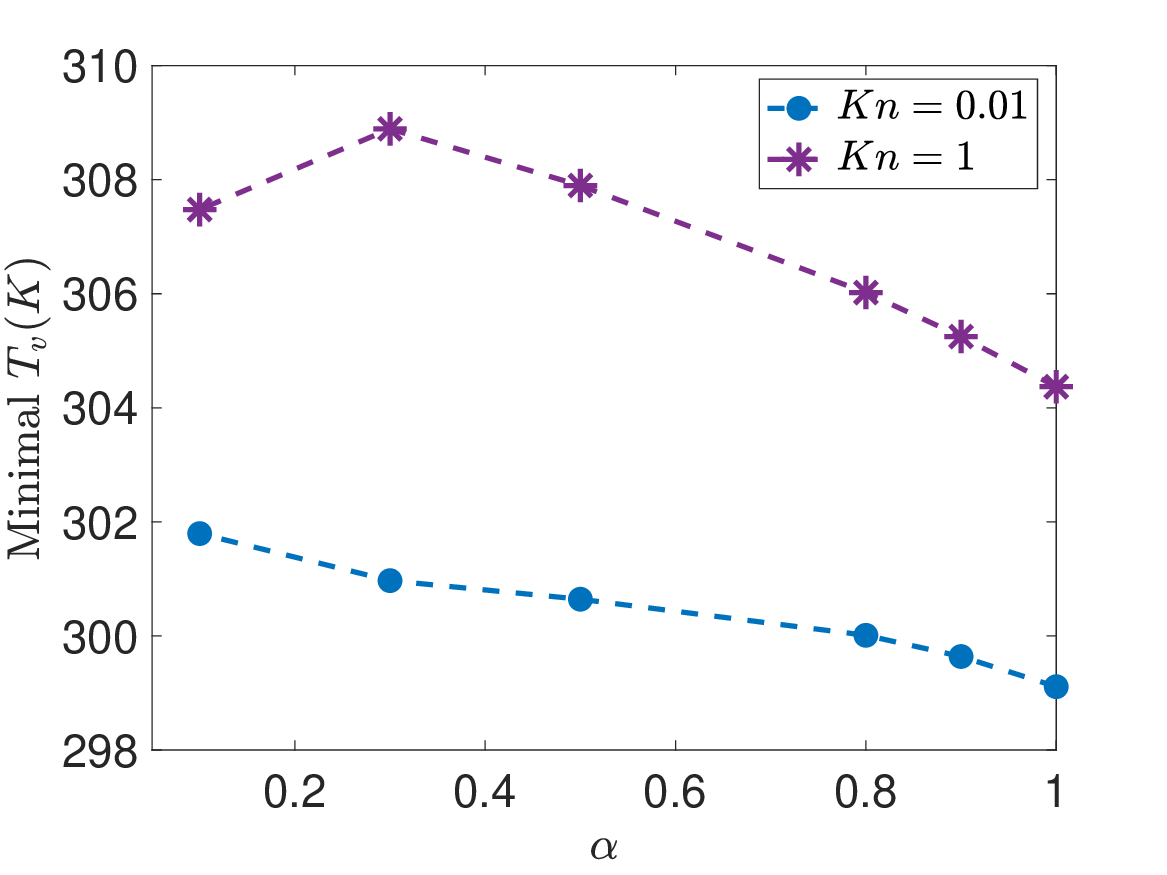}
    \caption{ }
    \label{fig:Tmin_change_a}
    \end{subfigure}
     \hfill
     \begin{subfigure}[b]{0.32\textwidth}
         \centering
         \includegraphics[width=\textwidth]{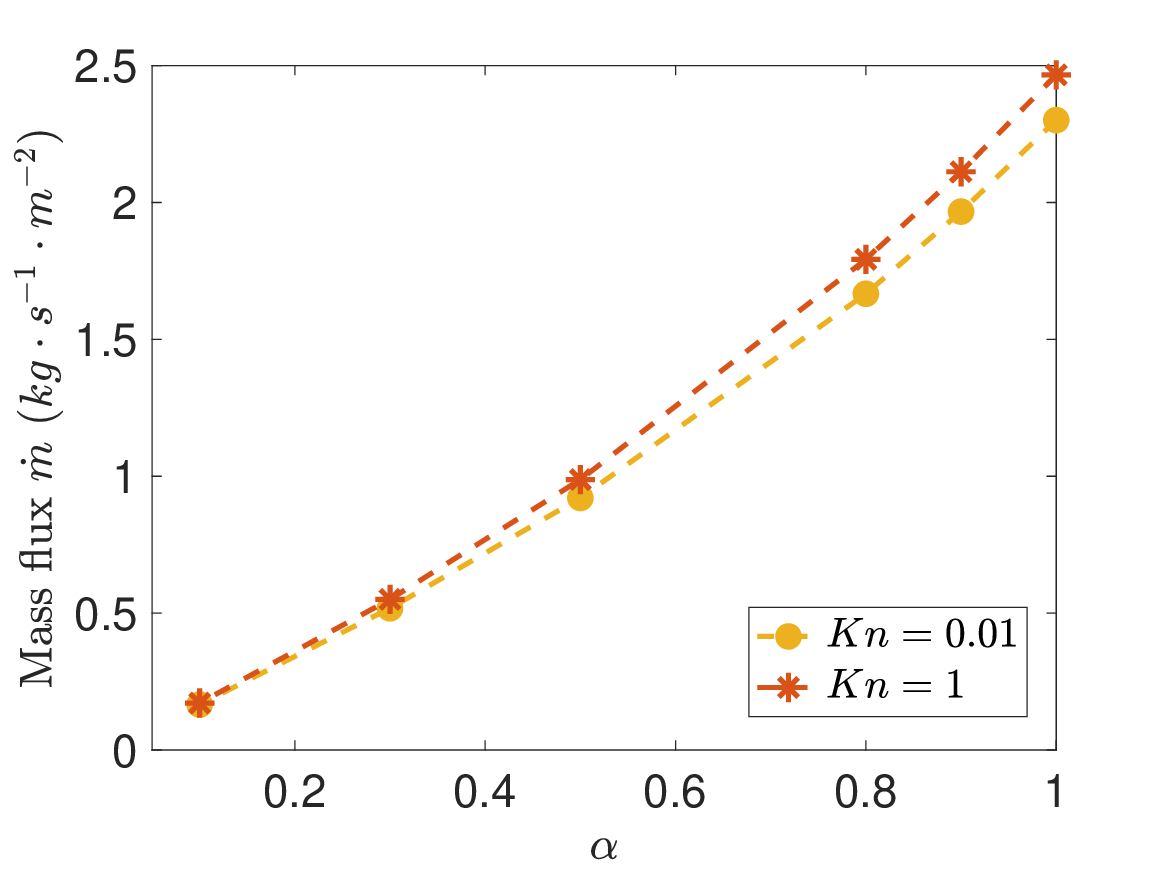}
    \caption{ }
    \label{fig:massf_change_a}
    \end{subfigure}
    \caption{\new{Dependence of Evaporative Refrigeration on Parameters. 
    (a) Impact of liquid film thickness on temperature distributions for fixed $(Kn=0.01, \alpha=0.5)$, showing thinner liquid film leads to lower vapor temperature.
    (b)\&(c) Impact of accommodation coefficient and Knudsen number (representing gap size) on vapor temperature distributions for fixed $d=0.1\mu m$, showing larger accommodation coefficient value and larger gap size (smaller $Kn$ number) leads to stronger cooling effect. 
    (d)\&(e) Minimal vapor temperatures and mass fluxes as a function of the liquid film thicknesses (d) and accommodation coefficient (e).
    (f) Mass fluxes as a function of the accommodation coefficients at two different $Kn$ numbers. Small gap size leads to higher mass flux.}}
    \label{fig: change d and change a}
\end{figure*}

In Fig.~\ref{fig: change d and change a}, we present the effects of different key parameters that impact the refrigeration effect, including the temperature distributions in the vapor gap, the minimum temperature of the vapor achieved, and the mass flow rates. For all the conditions, the temperature distributions are inverted. This is not surprising since the previous studies already showed that temperature inversions occur when 
\(\beta = \frac{T}{\rho}\frac{d\rho}{dT}=\frac{\Gamma}{RT_0}-1\geq \beta_c\), where $\Gamma$ is the latent heat and the critical value $\beta_c$ is about $3.5$~\cite{Pao-oppo-sign-1971, Cipolla-1974, SO-1978-kinetic, Chen-2023-parallel}. For \new{parameters used similar to} water, $\beta=18$, and hence the temperature distribution in between the gap is always inverted. 
\new{The Knudsen number is used to distinguish regimes. The small Knudsen number of $Kn=0.01$ represents that the vapor region can be treated as continuum, while $ Kn=1$ represents a transition regime where some molecules traverse the region without collisions and others undergo intermolecular collisions. We show results for three film thicknesses $d=1\mu\mathrm{m}, 0.5\mu\mathrm{m}, 0.1 \mu\mathrm{m}$ to illustrate the importance of the liquid film. Although these are very thin liquid films, they are encountered in boiling and evaporation in meniscus regions. The accommodation coefficient $\alpha=0.1$ is chosen to represent strong reflection and $\alpha=1$ for strong accommodation.}

It can be seen \new{from Fig.~\ref{fig: change d and change a}} that among the three parameters, the Knudsen number and the liquid film thickness have strong effects on the cooling. \new{While the effect of the accommodation coefficient had been studied before, this work is the first BTE-based study including the effect of the liquid layer thickness. The strong effect of the liquid layer thickness is less recognized.}

As the liquid film gets thinner, the minimal vapor temperature drops, as \new{shown} in Fig.~\ref{fig:Temp_change_d} \new{and further illustrated in Fig.~\ref{fig:Tmin_massf_change_d}. With thinner liquid film,} the heat transferred through the liquid layer is higher, leading to larger mass flux. \new{In Figure~\ref{fig:Temp_change_d}, the thinnest liquid film of $d=0.1\mu\mathrm{m}$ (dashed green line) leads to the lowest temperature in the vapor region.}
Larger accommodation coefficient values usually lead to larger mass flux [Fig.~\ref{fig:massf_change_a}] \new{and correspondingly larger heat flux, leading to lower minimal vapor temperature (Fig.~\ref{fig:Tmin_change_a})}. Fig.~\ref{fig:Temp_change_a_Kn001} and Fig.~\ref{fig:Temp_change_a_Kn1} show that a larger gap, i.e., a smaller Knudsen number, leads to lower vapor temperature, i.e., stronger refrigeration effect. \new{Comparing $Kn=0.01$ (Fig.~\ref{fig:Temp_change_a_Kn001}) and $Kn=1$ (Fig.~\ref{fig:Temp_change_a_Kn1}), we see that the vapor temperature drops below that of the cold wall only when $Kn=0.01$.} This is because at a smaller gap size, i.e., a larger Knudsen number, more heat is conducted in the reverse direction \new{due to the inverted temperature profile,} diminishing the refrigeration effect at the interface. \new{These results suggest that one can use a large vapor gap to probe the refrigeration effect, which makes the experiments easier. On the other hand, in applications such as membrane distillation, the reverse heat conduction may be useful in proving the efficiency~\cite{Chen-2023-parallel}. In addition, a smaller gap size leads to larger mass flux (Fig.~\ref{fig:massf_change_a}).}

\begin{figure*}
\begin{subfigure}[b]{0.32\textwidth}
         \centering
         \includegraphics[width=\textwidth]{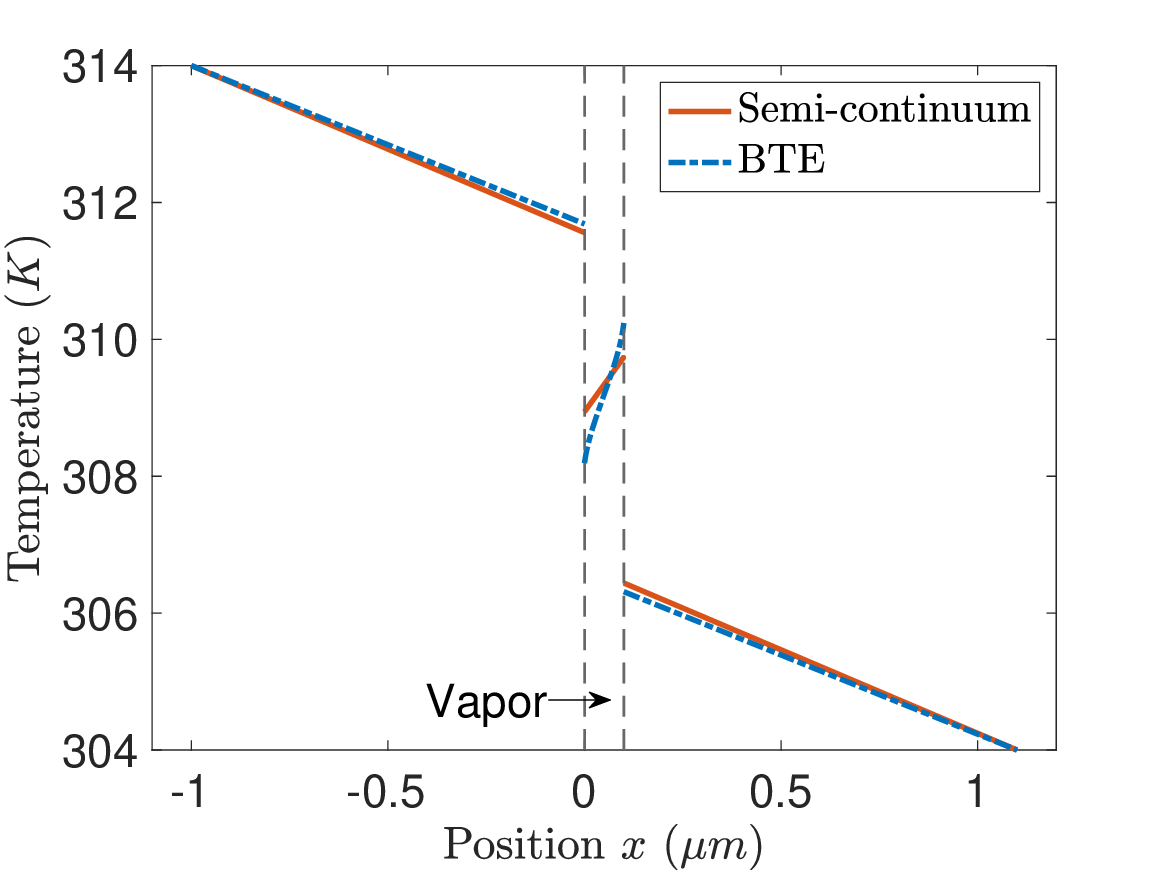}
    \caption{ }
    \label{fig:Temp_semi_Kn1}
     \end{subfigure}
     \hfill
     \begin{subfigure}[b]{0.32\textwidth}
         \centering
         \includegraphics[width=\textwidth]{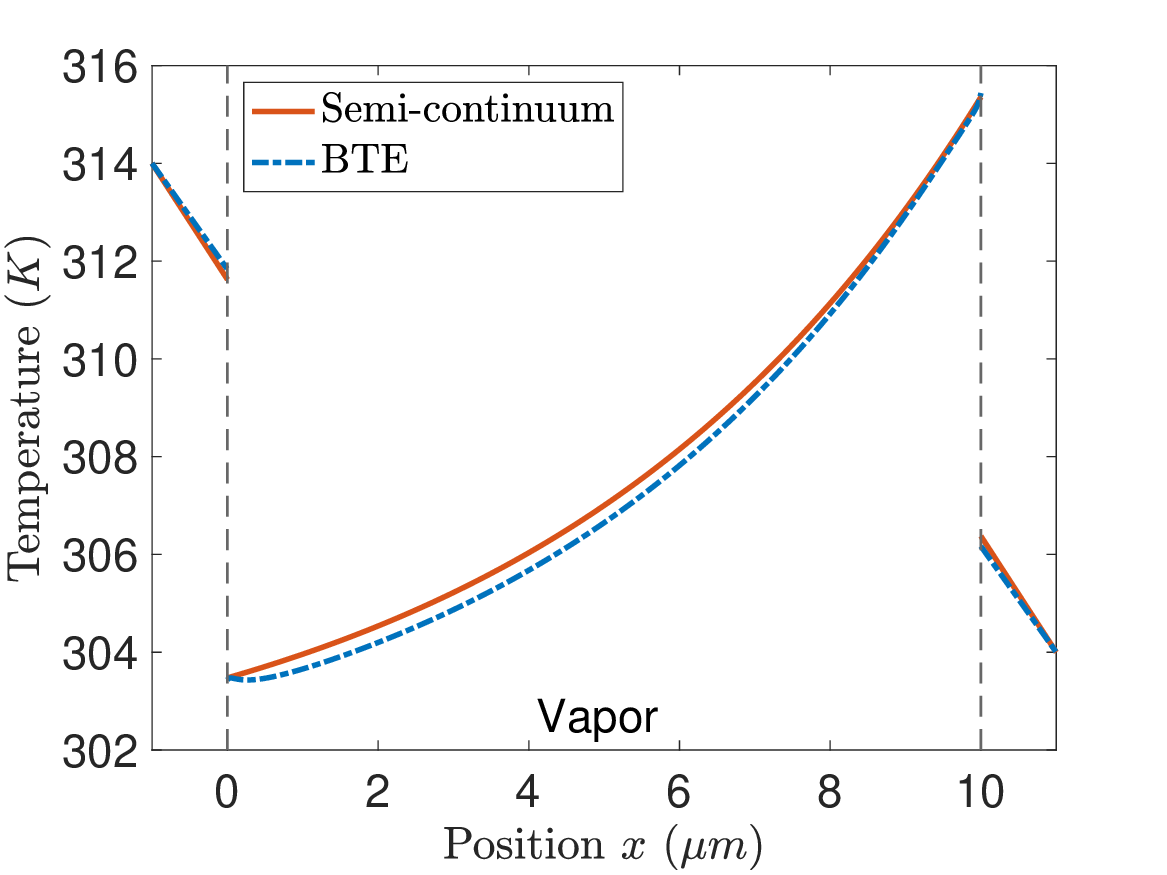}
    \caption{ }
    \label{fig:Temp_semi_Kn001}
    \end{subfigure}
    \hfill
     \begin{subfigure}[b]{0.32\textwidth}
         \centering
         \includegraphics[width=\textwidth]{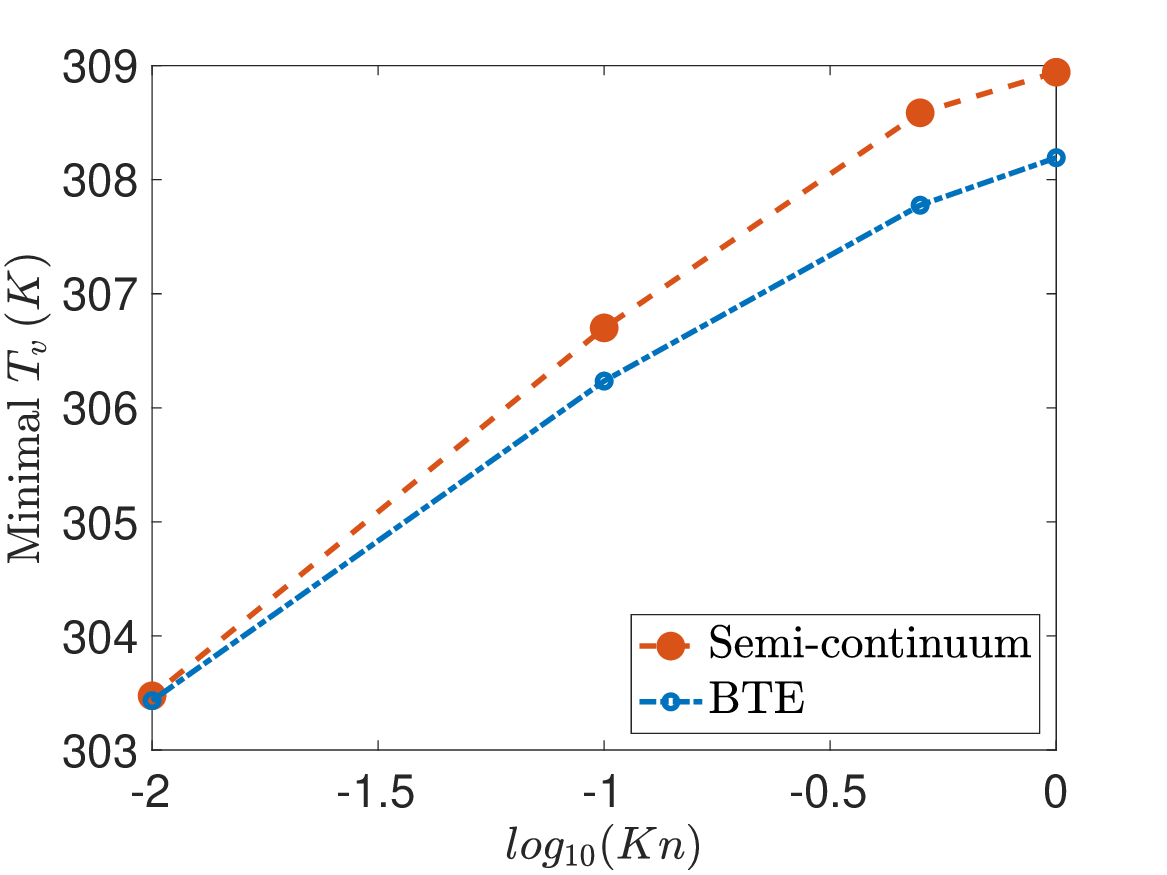}
    \caption{ }
    \label{fig:Tmin_change_Kn}
    \end{subfigure}
    \caption{\new{Comparison between semi-continuum solutions and BTE solutions. 
    (a)\&(b) Temperature distributions for two different Knudsen numbers. The vapor temperature obtained by the BTE solution has a lower vapor temperature and undergoes a bigger temperature change.
    (c) The minimum vapor temperature obtained by the semi-continuum solutions and the BTE solutions for different Knudsen numbers at the semi-log scale for fixed $(d=1\mu m, \alpha=0.5)$, showing better agreement at smaller $Kn$ numbers.}}
    \label{fig: change Kn}
\end{figure*}

The results obtained here enable us to compare with results obtained by Chen~\cite{Chen-2023-parallel} \new{using} the semi-continuum approach through coupling the interfacial mass and heat flux conditions to continuum formulations for both the liquid and the vapor regions, as shown in Fig.~\ref{fig:Temp_semi_Kn1} and Fig.~\ref{fig:Temp_semi_Kn001}. The semi-continuum approach predicts similar trends as the solution of the Boltzmann transport equation in terms of the temperature inversion and the discontinuities at the two interfaces. The semi-continuum approaches cannot predict the small temperature reversal inside the Knudsen region that exists in some cases as shown in Fig.~\ref{fig:Temp_dist_example}.  When $Kn\ll1$, the difference between the semi-continuum approach and that of the Boltzmann transport equation is small, \new{as seen in Fig.~\ref{fig:Temp_semi_Kn001} and Fig.~\ref{fig:Tmin_change_Kn}.}  However, at larger $Kn$ numbers (or equivalently, with smaller $L$), the semi-continuum approach \new{(the solid red line)} gives higher minimum vapor temperature values and thus underestimates the refrigeration effect as shown in Fig.~\ref{fig:Temp_semi_Kn1} \new{and Fig.~\ref{fig:Tmin_change_Kn}. As the Knudsen number grows, the minimal vapor temperature predicted by the semi-continuum solution is always above that by the BTE solution. Since the applied wall temperature difference is $10\Kelvin$, the $1\Kelvin$ discrepancy between the minimal temperature as shown in Fig.~\ref{fig:Tmin_change_Kn} represents $10\%$ difference.}

\section{\new{Conclusion}}

To summarize, we verify the existence of the evaporative refrigeration effect by solving the \new{BTE} coupled to liquid films for evaporation and condensation between two parallel plates. \new{Under certain conditions,} the vapor phase temperature in the evaporation side can drop below that of the condensing wall. This evaporative refrigeration effect is mainly due to the highly asymmetric molecular distribution function at the liquid-vapor interface, although the continued expansion in the Knudsen layer can lead to additional cooling.  Although the inverted temperature profile and temperature discontinuities have been studied before, our treatments include liquid films on both solid walls, making the predictions more realistic.  Our rigorous solution shows that the interfacial heat flux and mass flux boundary conditions developed before~\cite{GC-2022-flux-bc} are appropriate to treat such complicated problems using continuum approximation for both the liquid and the vapor phase. \new{The error of temperature between two approaches of the order of $10\%$ exists, and the errors of mass flux and heat flux between two approaches are of the order of $4\%$ at large Knudsen numbers. Our work shows that including the liquid film is important for the understanding. This study helps} point to directions in experimental validation, since larger gaps should be more favorable for experimental observations. 

\new{However, we would mention the limitations of our study and the insights provided for future work. In our study, although the thermal properties are chosen to be similar to those of water, only monatomic gas was considered. In this sense, it is the trend observed, rather than exact values, that is of interest. As past studies showed, including the polyatomic effect will not change the trend~\cite{CFF-1985-paradox}.}

\section*{Supplementary Material}
In the supplementary material, a detailed asymptotic calculation for small $Kn$ is provided, including the derivation of mass flux and energy flux in the bulk region and near the interfaces, as well as a review of the semi-continuum solution introduced by Chen~\cite{GC-2022-flux-bc}.

\section*{Acknowledgments}
P.C. and Q.L. acknowledge support from NSF-DMS-2308440. G.C. acknowledges support from MIT Bose Award and UMRP.

\section*{Conflict of Interest}
The authors have no conflicts to disclose.

\section*{Author Contributions}
Conceptualization: G.C.; Data curation: P.C., Q.L. and G.C.; Formal analysis: P.C., Q.L. and G.C. ; Investigation: P.C., Q.L. and G.C.; Methodology: G.C., Q.L. and P.C.; Software: P.C., Q.L., G.C.; Writing – original draft: G.C., Q.L. and P.C..

\section*{Data Availability Statement} The data that support the findings of this study are available from the corresponding author upon reasonable request.

\bibliographystyle{elsarticle-num} 
\bibliography{my_ref}

\end{document}